\documentclass[10pt,journal]{IEEEtran} 
\usepackage{subfigure}
\usepackage{times}
\usepackage{amsthm}
\theoremstyle{definition}

\long\def\@makecaption#1#2{\ifx\@captype\@IEEEtablestring%
\footnotesize\begin{center}{\normalfont\footnotesize #1}\\
{\normalfont\footnotesize\scshape #2}\end{center}%
\@IEEEtablecaptionsepspace
\else
\@IEEEfigurecaptionsepspace
\setbox\@tempboxa\hbox{\normalfont\footnotesize {#1.}~~ #2}%
\ifdim \wd\@tempboxa >\hsize%
\setbox\@tempboxa\hbox{\normalfont\footnotesize {#1.}~~ }%
\parbox[t]{\hsize}{\normalfont\footnotesize \noindent\unhbox\@tempboxa#2}%
\else
\hbox to\hsize{\normalfont\footnotesize\hfil\box\@tempboxa\hfil}\fi\fi}
\makeatother
\usepackage{booktabs}
\usepackage{lscape}
\usepackage{comment}
\usepackage{verbatim}
\usepackage{ltablex}
\usepackage[numbers]{natbib}
\usepackage{lscape}
\usepackage[ruled]{algorithm2e}
\usepackage{wrapfig}
\usepackage{color,soul}
\usepackage{longtable}
\usepackage{amssymb,amsmath}
\usepackage{multirow}
\usepackage{varwidth}
\usepackage{xcolor, colortbl}
\usepackage{xcolor}
\usepackage[pdftex]{graphicx}
\usepackage{stfloats}
\usepackage{graphicx}
\usepackage{tikz}
\usetikzlibrary{er}
\usetikzlibrary{shapes,snakes}
\usetikzlibrary{shapes.gates.logic.US,trees,positioning,arrows}
\usepackage{pgfplots}
\pgfplotsset{width=18cm, height=6cm}
\usepackage{pifont}
\newcommand{\cmark}{\text{\ding{51}}}
\newcommand{\xmark}{\text{\ding{55}}}
\usepackage[normalem]{ulem}
\usepackage[switch,columnwise]{lineno}
\usepackage{pgf-pie}
\SetAlFnt{\small}
\SetAlCapFnt{\small}
\SetAlCapNameFnt{\small}
\SetAlCapHSkip{0pt}
\IncMargin{-\parindent}
\newcolumntype{P}[1]{>{\centering\arraybackslash}p{#1}}

\newcommand{\new}{\textcolor{black}}

\usepackage{hyperref}
\usepackage{enumitem}
\pgfplotsset{compat=1.14}
\setlist[itemize]{leftmargin=*}

\begin{document}

\title{Toward Proactive, Adaptive Defense: \\ A Survey on Moving Target Defense}

\author{Jin-Hee~Cho, Dilli P. Sharma, Hooman Alavizadeh, Seunghyun Yoon, Noam Ben-Asher, Terrence J. Moore, Dong Seong Kim, Hyuk Lim, and Frederica F. Nelson
\thanks{Jin-Hee Cho is with The Virginia Tech, Falls Church, VA, USA, email: jicho@vt.edu}
\thanks{Dilli P. Sharma is with The University of Canterbury, Christchurch, New Zealand, email: dilli.sharma@pg.canterbury.ac.nz}
\thanks{Hooman Alavizadeh is with The Massey University, Auckland, New Zealand, email: h.alavizadeh@massey.ac.nz}
\thanks{Dong Seong Kim is with The University of Queensland, Australia, email: dan.kim@uq.edu.au}
\thanks{Seunghyun Yoon and Hyuk Lim are with The Gwangju Institute of Science and Technology, Gwangju, Republic of Korea, email: \{seunghyunyoon, hlim\}@gist.ac.kr}
\thanks{Noam Ben-Asher is with Boston Fusion, Brookline, MA, USA, email: noam.ben.asher@gmail.com}
\thanks{Terrence J. Moore, and F. F. Nelson are with US Army Research Laboratory, Adelphi, MD, USA, email: \{terrence.j.moore, frederica.f.nelson\}.civ@mail.mil}
\thanks{This work has been submitted to the IEEE for possible publication. Copyright may be transferred without notice, after which this version may no longer be accessible.}
}

\newpage

\maketitle

\begin{abstract}
Reactive defense mechanisms, such as intrusion detection systems, have made significant efforts to secure a system or network for the last several decades. However, the nature of reactive security mechanisms has limitations because potential attackers cannot be prevented in advance. We are facing a reality with the proliferation of persistent, advanced, intelligent attacks while defenders are often way behind attackers in taking appropriate actions to thwart potential attackers. The concept of moving target defense (MTD) has emerged as a proactive defense mechanism aiming to prevent attacks. In this work, we conducted a comprehensive, in-depth survey to discuss the following aspects of MTD: key roles, design principles, classifications, common attacks, key methodologies, important algorithms, metrics, evaluation methods, and application domains. We discuss the pros and cons of all aspects of MTD surveyed in this work. Lastly, we highlight insights and lessons learned from this study and suggest future work directions.  The aim of this paper is to provide the overall trends of MTD research in terms of critical aspects of defense systems for researchers who seek for developing proactive, adaptive MTD mechanisms.
\end{abstract}

\begin{IEEEkeywords}
Moving target defense, proactive defense, cybersecurity, attack surface, shuffling, diversity, redundancy
\end{IEEEkeywords}

\IEEEpeerreviewmaketitle

\section{Introduction} \label{sec:intro}

\subsection{Motivation} \label{subsec:motivation}
In the classic ``shell game,'' dating at least to the times of ancient Greece, a pea or ball is hidden under one of three shells or cups, and players gamble to guess the location of the pea after the shells have been moved. This game is also known as ``thimblerig,'' ``three shells and a pea,'' and the ``old army game'' and has variant forms, such as ``Three-card Monte.'' The advantage for the operator of the game is that sufficient movement of the shells will confuse the players of the location of the pea. This underlying idea exemplifies the philosophy of the so-called ``Moving Target Defense (MTD)'' strategy in cybersecurity~\cite{dhs}.

The basic axiom of MTD is that it is impossible to provide complete and perfect security for a given system. Given this, the objective is to enable normal functioning of the system (i.e., the normal provision of services) even in the presence of malicious actors seeking to compromise the system. Since attacks cannot be prevented, the goal is to defend against and thwart attacks. In practice, this goal is achieved via MTD by the manipulation of multiple system configurations to modify and control the attack surface, where the attacker engages with the system. MTD aims to increase uncertainty and complexity for any attacker of the system, to decrease the opportunities for the attacker to identify targets (e.g., vulnerable system components), and to introduce higher cost in launching attacks or scans (e.g., reconnaissance attacks). The desired result is that the attacker will waste time and effort without gaining useful intelligence about the system~\cite{hong2016assessing, dhs}.

Origins of the conception of MTD can be found under different names in the computer security literature. Some of these research areas include fault tolerance (or reliability using redundancy) even since the 1970's (e.g., $n$-version programming (NVP)~\cite{Avizienis85, Chen78}), reconfigurable computing (e.g., reconfigurable software~\cite{Compton02}, network reconfiguration~\cite{Baran89}), and/or bio-inspired cybersecurity (e.g., software / network diversity for cybersecurity~\cite{Knight16, Lala09}). In 2009, the Networking and Information Technology Research and Development (NITRD) Program explicitly emphasized the concept of MTD in terms of its effectiveness and efficiency, which can leverage existing resources~\cite{Ghosh09}. Since then, a fast-growing research community has formed with a focus on MTD, and many initial works have received significant attention due to its fascinating approach and merit. However, there has been relatively less effort to comprehensively understand the overall trends in terms of MTD classifications, attacks handled by MTD, metrics used to measure the performance of MTD techniques, and associated limitations of the existing MTD technologies. An overall in-depth understanding of MTD techniques will guide us to identify what aspects lack in existing MTD approaches and accordingly what directions should be pursued for future research. This is the aim of our survey paper.

\subsection{Comparison of Our Work and Existing MTD Survey Papers}
As significant attention has been paid to this emerging field of research in the past ten years, some efforts have been made to understand the state-of-the-art MTD technologies. \new{We compared our survey paper and the existing survey papers in terms of their contributions, key design, and classification together with the following principal questions:
\begin{itemize}
\item[] {\bf Q1}: What attacks are considered more/less likely to be handled by existing MTD techniques?
\item[] {\bf Q2}: What MTD techniques are more commonly explored?
\item[] {\bf Q3}: What theoretical and empirical methodologies are used in developing and/or evaluating MTD techniques?
\item[] {\bf Q4}: What specific limitations are identified in using a particular technique?
\item[] {\bf Q5}: What metrics are unavailable to measure the key effectiveness of MTD techniques in terms of system security and performance?
\end{itemize}
We discussed the contributions of each survey paper on MTD based on the main criteria and principle questions in Table~\ref{table:Surveys}.}
\begin{table*}[t]
\caption{\new{Comparison of the Key Contributions of Our Survey Paper Compared to Other Existing Survey Papers.}}
\vspace{-2mm}
\label{table:Surveys}
\centering
\begin{tabular}{@{}lllllll@{}}
\toprule
\textbf{Criteria}                                                                                                                 & \textbf{Our Survey}                                                                       & \begin{tabular}[c]{@{}l@{}} Cai et al. \\(2016) \cite{Cai16}\end{tabular}& \begin{tabular}[c]{@{}l@{}}Ward et al. \\ (2018) \cite{Ward18-survey}\end{tabular} & \begin{tabular}[c]{@{}l@{}}Lei et al. \\ (2018) \cite{Lei18-survey}\end{tabular} & \begin{tabular}[c]{@{}l@{}}Zheng and Namin \\(2019) \cite{Zheng:SurveyOnMTD2019}\end{tabular} & \begin{tabular}[c]{@{}l@{}}Sengupta et al. \\(2019) \cite{sengupta2019survey}\end{tabular} \\ \midrule
Principle questions not addressed (Q1-Q5)                                                                                 &                                                                               &  Q1, Q4, Q5                                                                  & Q1-Q3, Q5                                                               & Q5                                                                         & Q5                                                                                       & Q4                                                                                    \\ \midrule
\multicolumn{7}{c}{\textbf{Categorization/Classification in the Survey} (Classified and Surveyed (\cmark) - Surveyed but not Classified (\textit{S}-only) - Neither Classifies nor Surveyed (\xmark))}                                                                                                                                                                                                                                                                                                                                                                                                                                                       \\ \midrule
Evaluation of effectiveness and efficiency of MTD                                                                                 & \cmark                                                                              & \cmark                                                                 & \cmark                                                        & \cmark                                                               & Partially                                                                                & \cmark                                                                          \\
\begin{tabular}[c]{@{}l@{}}Key Design Principles \\ ~(WHAT to move, WHEN to move, HOW to move)\end{tabular}                        & \cmark                                                                              & \cmark                                                                        & \textit{S}-only                                                       & \xmark                                                                   & \cmark                                                                             & \cmark                                                                          \\
\begin{tabular}[c]{@{}l@{}}Application Domain\\  ~(Cloud, IoT, etc.)\end{tabular}                                                  & \begin{tabular}[c]{@{}l@{}}\scriptsize Cloud, SDNs, IoT, \\ CPSs, Enterprise\end{tabular} & \textit{S}-only                                                                       & \textit{S}-only                                                              & \begin{tabular}[c]{@{}l@{}}\scriptsize Cloud, SDNs,\\ IoT\end{tabular}     & \begin{tabular}[c]{@{}l@{}}\scriptsize Cloud, SDNs,\\ IoT\end{tabular}                                                             & \cmark                                                                          \\
\begin{tabular}[c]{@{}l@{}}Combination of MTD techniques vs \\ ~different layers (as Table~\ref{table-movingelements})\end{tabular} &  \cmark                                                                             & \xmark                                                                       & \xmark                                                              & \xmark                                                                     & \xmark                                                                                   & \xmark                                                                                \\
\begin{tabular}[c]{@{}l@{}}Types of Attack behaviors\\ ~(such as attacker knowledge: Intelligent or not)\end{tabular}              & \cmark                                                                              & \textit{S}-only                                                                      & \xmark                                                               & \cmark                                                                 & \begin{tabular}[c]{@{}l@{}}Partially\\ \scriptsize (Only attack types)\end{tabular}      & \cmark                                                                            \\
Types of evaluation methods                                                                                                       & \cmark                                                                                & \cmark                                                                       & \textit{S}-only                                                          & \cmark                                                                 & \xmark                                                                                   & \cmark                                                                            \\
\begin{tabular}[c]{@{}l@{}}Discussing Key concept of MTD \\ ~(deception vs. MTD)\end{tabular}                                      & \cmark                                                                              & Partially                                                                       & \xmark                                                           & Partially                                                                  & \cmark                                                                               & Partially                                                                             \\ \midrule
\multicolumn{7}{c}{\textbf{Discussing Limitation, Pros and Cons of MTD Techniques}}                                                                                                                                                                                                                                                                                                                                                                                                                                                                                                                                                                \\ \midrule
\begin{tabular}[c]{@{}l@{}}Discussing Key Limitations/\\ Pros and Cons of Techniques\end{tabular}                                 & \cmark                                                                              & \xmark                                                                    & Partially                                                              & \xmark                                                                     & \xmark                                                                                   & Partially                                                                             \\
\begin{tabular}[c]{@{}l@{}}Discussion for insights and future directions\\ of MTD literature\end{tabular}                         & Comprehensive                                                                             & Partially                                                                     &  \xmark                                                           & \cmark                                                                 & Partially                                                                                & \cmark                                                                            \\ \bottomrule
\end{tabular}
\end{table*}

\new{
MIT Lincoln Lab published a technical report providing an extensive survey on MTD in 2013~\cite{okhravi2013survey}, with a substantially expanded and updated edition five years later~\cite{Ward18-survey}. Both surveys are very helpful for readers interested in learning each MTD technique with the classification based on `what to move' in deploying MTD. Specifically, they classified MTD techniques under five categories: dynamic data, dynamic software, dynamic runtime environments, dynamic platforms, and dynamic networks. The second edition~\cite{Ward18-survey} particularly includes an extensive list of MTD techniques for each category. Consequently, they provided the types of attacks based on the classified MTD. In our survey, we also categorize the MTD based on techniques and operational layers which covers the taxonomies used by~\cite{Ward18-survey} as well as provides the ways in which those techniques can be combined together. The authors described each technique under these criteria, which is useful as a manual description of the techniques. However, this kind of survey does not provide the succinct overall summary and trends of the MTD research field that would enable the readers to discover answers to the principal questions (e.g., Q1-Q5).  On the other hand, our survey paper aims to answer these questions.}

\new{\citet{farris2015quantification} studied the existing MTD techniques to quantify their cost and effectiveness through an expert opinion survey.  They reached 120 cyber defense experts and evaluated 39 MTD techniques. They ranked the most seven dominant ones based on the quantification criteria in terms of cost against effectiveness.  They found that the best MTD techniques in terms of operating costs and the overall effectiveness are the techniques using address space layout or instruction set randomization while the worst is one performing network randomization.  Although their evaluation for quantifying costs are well-evaluated based on the implementation and performance costs, their consideration for quantifying the effectiveness of MTD techniques is limited to the attacker's workload. They do not explicitly define what the effectiveness is in terms of other important security metrics. Moreover, a set of studied literature is limited to only 39 papers in which only 23 of them are opted as dominant studies and used for the survey.}

\new{
\citet{Cai16} also conducted an extensive survey on the state-of-the-art MTD research by describing theoretical approaches, strategies based on key characteristics, and evaluation methods used in those existing MTD approaches.  The authors provided a function-and-movement model categorizing MTD techniques in a systematic way in terms of (1) three layers of implementation including software, running platform, and physical network; (2) function and movement models; and (3) MTD, its strategies, and its evaluation. However, from the reader's perspective, it is challenging to understand the principle questions Q1 and Q4-Q5. Although this classification covers a large portion of literature, it misses other perspectives of MTD such as testbeds, a combinations of MTD techniques, attack types, and the aforementioned limitations. In our survey paper, these three questions are answered in Sections~\ref{sec:mtd_classification}, \ref{sec:threat-models}, and \ref{sec:metrics}, respectively.}

\new{
Recently, \citet{Lei18-survey} published a survey paper on MTD techniques.  Although the authors conducted extensive surveys of the existing MTD techniques in terms of strategy generation, shuffling implementation, and performance evaluation, this work missed many MTD taxonomies to categorize and understand the existing MTD techniques. This work also did not survey an in-depth security analysis of MTD techniques based on well-known security metrics used to evaluate the current state-of-the-art MTD approaches. The MTD classification used in \cite{Lei18-survey} is based on multiple criteria, such as the theoretical basis (e.g., game theory, machine learning, genetic algorithms), the techniques (e.g., transformation or shuffling of system configurations), and the purpose of existing MTD approaches, which lack the consistent perspective to classify and analyze the existing MTD works.  Due to this reason, one may need to have a more consistent classification to clearly understand the overall trends of MTD research and easily obtain insights from the existing MTD methodologies.}

\new{~\citet{Zheng:SurveyOnMTD2019} published a survey paper on architectural aspects of the MTD techniques based on application, operating system, and network levels. They classified various types of MTD techniques, including software diversification via middle-ware, address space layout randomization, instruction set randomization, IP randomization, virtualization, decoys, and software-defined or lightweight based MTDs, at different levels.  However, they did not clearly distinguish MTD techniques from MTD architectures. For instance, software diversification, address space layout randomization, or IP randomization are MTD techniques while middle-ware based software diversification and software-defined MTD are the MTD deployment architectures. In addition, they did not discuss targeted attackers characteristics, evaluation methods, and suitable metrics of the existing MTD techniques, which we addressed in this survey paper.}

\new{Very recently,~\citet{sengupta2019survey} conducted a substantial survey to study and categorize the proposed defensive MTD tools, techniques, and strategies. The authors categorized MTD approaches in a structured way and introduced MTD taxonomies based on the following technique formalism: designing key movements, implementation of MTD (e.g., testbeds), and evaluation of MTD effectiveness (e.g., qualitative and quantities metrics).
However, some critical aspects of the implementation layers and combinations of MTD techniques were missing in their classification.
On the other hand, our survey paper here classified MTD techniques based on Table~\ref{table-movingelements} consisting of MTD techniques categorization and the layer of implementation. } 

\new{The existing MTD survey papers~\cite{Cai16, Lei18-survey, okhravi2013survey, sengupta2019survey, Ward18-survey, Zheng:SurveyOnMTD2019} have provided different perspectives of MTD classifications and highly useful information overall. However, they missed some important information that may be highly needed for researchers who want to start conducting MTD research. The common examples may include measurement methods, metrics, common attackers countermeasured by MTD, distinctions between MTD and other defense mechanisms (e.g., defensive deception, intrusion detection) along with the discussions of pros and cons for each MTD techniques and MTD specific for different system environments, which we tried to cover all in our survey paper. To effectively demonstrate the commonalities and differences between our survey paper and the existing survey papers, we created a table summarizing their comparisons in Table~\ref{table:Surveys}.}

Furthermore, strictly speaking, MTD is different from deception in that it does not intentionally provide false information to deceive an attacker while deception does, even if these two approaches may lead to a similar outcome, such as confusing the attacker and misleading the attacker's decision making process. All the existing survey papers discussed above~\cite{Cai16, Lei18-survey, okhravi2013survey, Ward18-survey} treat deception techniques as a subset of MTD techniques. On the other hand, the deception research community treats MTD as a subset of deception techniques~\cite{Pawlick17-game-deception}. Often times, deception is combined with MTD in which deception is used to change the attack surface. 
Hence, in our survey paper, we take deception techniques (e.g., decoy nodes, honeypots / honeynets, fake information dissemination on system configuration) as the part of MTD techniques only when they are used to change the attack surface, which is the key concept of MTD. We will discuss more details of the differences between these two techniques in Section~\ref{subsec:deception_mtd}.

\subsection{Key Contributions \& Scope} \label{subsec:scope_contribution}
Unlike the existing survey papers on MTD, our survey paper has the following {\bf novel key contributions}: 
\begin{itemize}
\item We extensively surveyed the state-of-the-art MTD technologies and classified them based on three operation types: shuffling, diversity, and redundancy (SDR). This classification was first defined in our prior work~\cite{hong2016assessing} but has not been used to extensively survey the existing MTD techniques. Hence, this work is the first to examine the overall trends of existing MTD techniques based on this classification type. We chose this classification type because it can embrace MTD techniques across different systems (or networks) layers as long as they have the common goal(s) to achieve. We discuss more about the operation-based MTD classification in Section~\ref{subsec:goal-mtd-classification}.
\item We investigated what types of attack behaviors are addressed by the state-of-the-art MTD techniques. In particular, we discussed them in the context of the operation-based classification (i.e,, shuffling, diversity, and redundancy). In addition, we discussed the limitations of current attack models.
\item We discussed the key distinct contributions of MTD technologies compared to conventional security counterparts. We discussed the differences and commonalities between MTD and cyberdeception, whose concepts are often confused \new{as being an MTD, since the distinction between the two has not been clarified in the literature.}
\item We surveyed what specific methodologies or algorithms are leveraged to develop MTD techniques, which embrace three major theoretical backgrounds of those MTD approaches, including game theory, genetic algorithms, and machine learning.
\item We also surveyed various types of evaluation methods used to validate the performance of MTD techniques, including analytical models, simulation, emulation, and real 
\new{testbed environments.}
\item We surveyed metrics used to assess the quality (i.e, performance and security) of MTD techniques considering the measures in terms of the perspective of attackers or defenders.
\item We examined how MTD technologies have been applied in various application domains, including enterprise networks, Internet-of-Things (IoT), cyber-physical systems (CPSs), software defined networks (SDNs), and cloud computing.
\item Lastly, we discussed insights and limitations obtained from this extensive survey and suggested future work directions based on lessons learned from this study.
\end{itemize}

\subsection{Paper Structure}
This paper is organized as follows:
\begin{itemize}
\item Section~\ref{sec:roles_compo} discusses the key roles and design principles of MTD techniques.
\item Section~\ref{sec:mtd_classification} discusses the existing MTD classifications types.
\item Section~\ref{sec:mtd_techniques} discusses MTD techniques under the classification of operation-based MTD consisting of shuffling, diversity, and redundancy. Additionally, we discuss existing hybrid approaches using the mixture of these techniques.
\item Section~\ref{sec:threat-models} describes attack behaviors handled by existing MTD approaches.  In addition, we address the limitations of current attack models considered by the existing MTD approaches.
\item Section~\ref{sec:modeling-solution-mtd} provides the existing MTD techniques developed by the following theoretical approaches: (1) game theory; (2) genetic algorithms; and (3) machine learning.
\item Section~\ref{sec:metrics} surveys metrics used to evaluate the effectiveness and efficiency of MTD approaches in terms of perspectives of either attackers or defenders.
\item Section~\ref{sec: evaluation-methods} gives the overall description of evaluation methods used to validate existing MTD techniques, including: (1) analytical models; (2) simulation models; (3) emulation models; and (4) real testbeds.
\item Section~\ref{sec:application-domains} provides the overall trends of how MTD techniques are applied under different application domains including: (1) enterprise networks; (2) Internet-of-Things (IoT); (3) cyber-physical systems (CPSs); (4) software defined networks (SDNs); and (5) cloud-based web services.
\item Section~\ref{sec:limitations} discusses the limitations of the existing MTD approaches covered in this work.
\item Section~\ref{sec:insights-future-work} discusses the insights and lessons learned from our study and suggests future research directions in the area of MTD research.
\end{itemize}

\section{Key Roles \& Design Principles of MTD} \label{sec:roles_compo}

\new{In this section, we provide backgrounds and key design principles of MTD. We discuss the two major roles of MTD in terms of the intrusion prevention mechanism and the detection of potential attackers. In addition, we clarified the design principles of MTD with three key questions, and summarized benefits and caveats of applying MTD techniques.}

\subsection{Key Roles} \label{subsec:mtd_role}
MTD can contribute to enhancing the security provided by an intrusion defense system \new{(IDS)} in two ways. First, MTD acts as an {\em intrusion prevention mechanism} by reducing the risk of potential intrusions to the system. Many MTD techniques, including Internet Protocol (IP) mutation, network topology changes, platform migration, and software stack shuffling, \new{change} 
the attack surface, thereby minimizing the potential of attackers in their discovery of vulnerable system components and in their attempts to \new{penetrate} 
into the system. 

Second, MTD can assist in the detection of potential attackers. Since MTD techniques can increase the complexity of an attack operation to identify a target, the intelligence about attack patterns or behaviors of potential attackers can be easily obtained by monitoring active activities (e.g., scanning) by the potential attackers. 
\new{This allows MTD techniques, categorized as intrusion protection mechanisms, to also indirectly assist an intrusion detection system to improve the detection of potential attacks.}

\begin{table*}
\caption{Comparison between the conventional defense vs. MTD} \label{table:comparison-mtd-conventional}
\vspace{-4mm}
\begin{center}
\begin{tabular}{|P{1.8cm}|P{7cm}|P{7cm}|}
\hline
 & {\bf Conventional Defense} & {\bf MTD} \\
\hline
{\bf Philosophy} & Building a secure system by covering all vulnerabilities is the key for security & It is impossible to build a perfectly secure system \\
\hline
{\bf Goal} & Make a system perfectly secure & Thwart attackers by changing attack surface \\
\hline
{\bf Defense Type} & Static defense using static system configurations fortifying assets and hardening defense systems & Dynamic defense using dynamic system configurations to build a agile system \\
\hline
{\bf Key Concern} & How to identify all potential system security vulnerabilities and eliminate them & How to optimally execute MTD operations to continuously change attack surface while meeting multiple system objectives concerning security and performance \\
\hline
{\bf Challenges} & Due to limited capability to detect all exhaustive, possible vulnerabilities and their high complexity, it is unfeasible to identify all possible vulnerabilities  & Optimally executing MTD operations is not a trivial task because frequently executing MTD operations is costly and often hurts service availability to users \\
\hline
\end{tabular}
\end{center}
\end{table*}
As Table~\ref{table:comparison-mtd-conventional} describes, compared to the characteristics of the conventional security techniques, MTD aims to provide {\em affordable, service-oriented defense} in terms of minimizing defense cost (e.g., communication or computation overhead), maximizing service availability to users, and meeting a required level of security. In particular, the underlying idea of MTD lies in `affordable defense' because it focuses on rearranging system configurations to confuse attackers while still employing existing security mechanisms. This means that MTD does not require \new{the development of}
new security mechanisms, which often need to go through tough, high-standard security analysis and accordingly may require \new{an excessive amount of time or effort.}
Therefore, MTD provides a new way of defense that 
\new{continues to leverage legacy resources but enhances security by dynamically changing the attack surface,} 
such as any aspects of system components (e.g., data, software, platforms, runtime environments, networks) to increase uncertainty and confusion for attackers.

\begin{figure*}[th!]
	\centering
	\includegraphics[height=5cm]{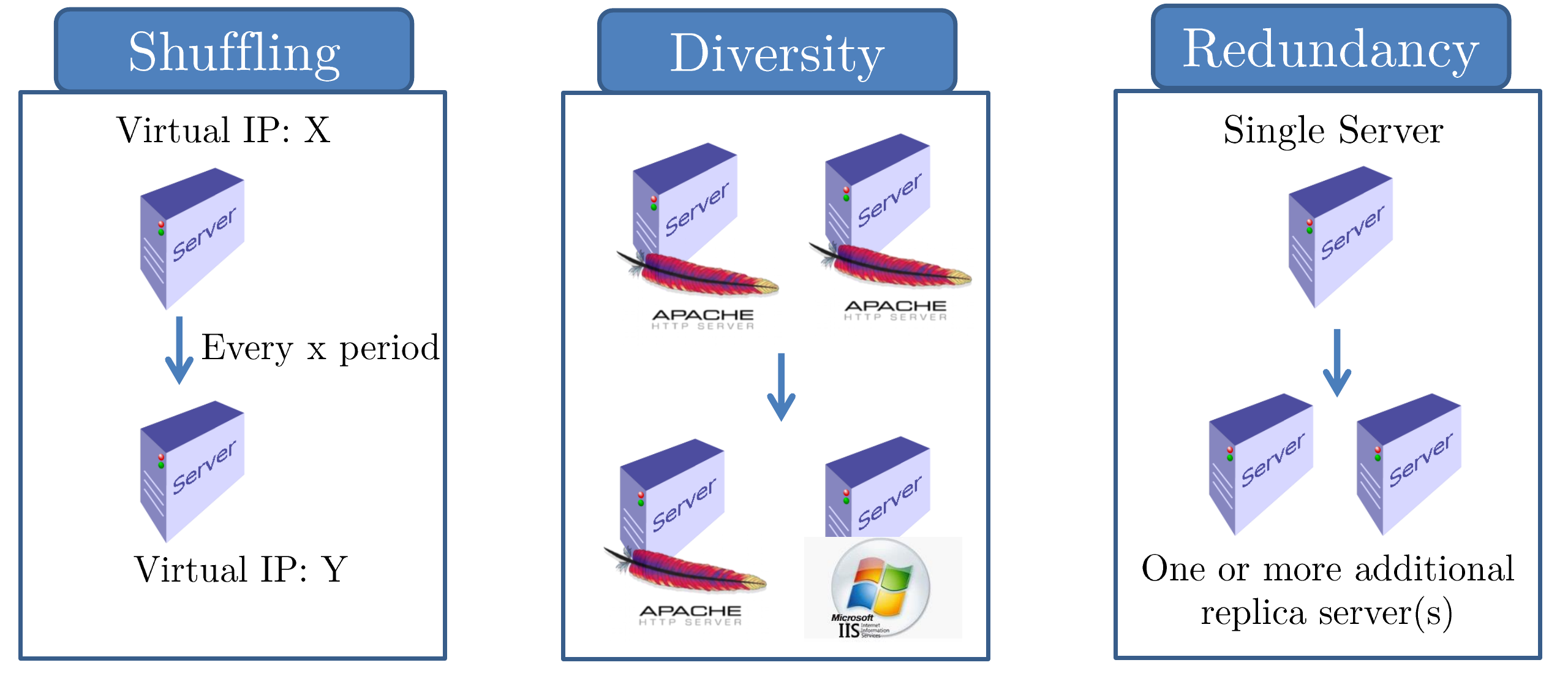}
	\caption{Examples of MTD techniques based shuffling, diversity, or redundancy.}
	\label{fig:how_to_move}
\end{figure*}

\begin{figure}[th!]
	\centering
	\includegraphics[width=0.45\textwidth, height=4cm]{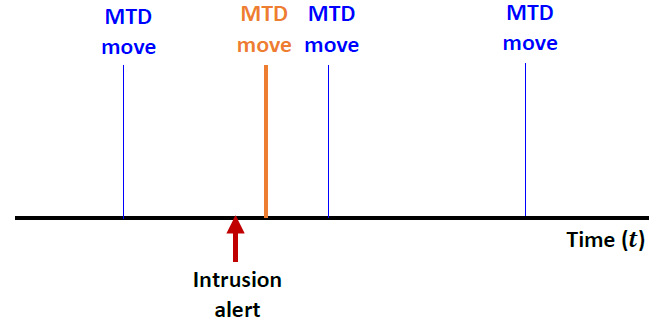}
	\caption{An example of an adaptive MTD: A blue vertical line represents a proactive MTD which is triggered at fixed MTD interval whereas a red line presents a reactive MTD which is triggered on an event (e.g, an intrusion detection alert).}
	\label{fig:when_to_move}
\end{figure}

\subsection{Key Design Principles} \label{subsec:mtd_decisions}
MTD was introduced as a proactive defense mechanism to prevent cyberattacks by continuously and dynamically changing the attack surface of systems~\cite{Ghosh09}. The fundamental design principle for developing MTD techniques lies in the decisions for the following three key questions~\cite{Cai16}: 
{\em what to move}, {\em how to move}, and {\em when to move}.

\begin{table*}[ht]
	\centering
	\caption{Moving elements of MTD techniques at different layers}
	\label{table-movingelements}
	\vspace{-2mm}
	\resizebox{\textwidth}{!}{%
		\begin{tabular}{|c|c|c|c|}
			\hline
			\rowcolor[HTML]{EFEFEF} 
			\cellcolor[HTML]{FFFFFF} & \multicolumn{3}{c|}{\textbf{MTD Techniques}} \\ \cline{2-4} 
			\multirow{-2}{*}{\textbf{Layers}} & \multicolumn{1}{c|}{\textbf{Shuffling}} & \multicolumn{1}{c|}{\textbf{Diversity}} & \multicolumn{1}{c|}{\textbf{Redundancy}} \\ \hline
			\cellcolor[HTML]{EFEFEF} & & Web: Appache, IIS, GWS etc. & Web service replica   \\ \cline{3-4}
			\cellcolor[HTML]{EFEFEF} &  & App: .Net Framework, Java, PHP etc. & Application replica \\ \cline{3-4}
			\cellcolor[HTML]{EFEFEF} & & Database: SQL server, MySQL, Oracle etc. & Database backup and replica  \\ \cline{3-4}
			\multirow{-4}{*}{\cellcolor[HTML]{EFEFEF}\textbf{Application}} & \multirow{-4}{*}{TCP/UDP port numbers}  & Others: Mail-server, Proxy-server etc. & Other service replica  \\ \hline
			\cellcolor[HTML]{EFEFEF} & & Windows: Windows server 2003/2008, Windows 9.x, 8, 10 etc. &    \\ \cline{3-3}
			\cellcolor[HTML]{EFEFEF} & & Linux: Redhat, Debian, Caldera etc. &  \\ \cline{3-3}
			\cellcolor[HTML]{EFEFEF} & & Solaris &   \\ \cline{3-3}
			\multirow{-4}{*}{\cellcolor[HTML]{EFEFEF}\textbf{OS-Host}} & \multirow{-4}{*}{IP address} & Others: Unix, HP-UX etc. &  \\ \cline{2-2} \cline{3-3} 
			\cellcolor[HTML]{EFEFEF}\textbf{VM-Instance} & Virtual IP address & Same as OS & \multirow{-5}{*}{Host OS and virtual machine  replica}  \\ \hline
			\cellcolor[HTML]{EFEFEF} & & Xen &  \\ \cline{3-3}
			\cellcolor[HTML]{EFEFEF} & & Vmware &   \\ \cline{3-3}
			\cellcolor[HTML]{EFEFEF} & & ESXi &   \\ \cline{3-3}
			\multirow{-4}{*}{\cellcolor[HTML]{EFEFEF}\textbf{Virtual Machine Manager (VMM)}} & \multirow{-4}{*}{Failover, Switchover} & Others: Kernel-based VM (KVM), Virtual-box (Vbox), IBM vSphere & \multirow{-4}{*}{Hypervisor's replica}   \\ \hline
			\multicolumn{1}{|l|}{\cellcolor[HTML]{EFEFEF}} & & Intel &   \\ \cline{3-3}
			\multicolumn{1}{|l|}{\cellcolor[HTML]{EFEFEF}} & & HP &    \\ \cline{3-3}
			\multicolumn{1}{|l|}{\cellcolor[HTML]{EFEFEF}} & & Sun Solaris &    \\ \cline{3-3}
			\cellcolor[HTML]{EFEFEF} \multirow{-4}{*}{\bf Hardware} & \multirow{-4}{*}{Hardware replacement}&  Others: ARM, Atmega & \multirow{-4}{*}{Hardware backups and replica} \\ \hline
		\end{tabular}%
	}
\end{table*}

\begin{itemize}
\item {\bf What to move}: `What to move' refers to what system configuration attribute (i.e., attack surface) can be dynamically changed to confuse attackers. The example system or network attributes that can be changed include instruction sets~\cite{Kc:ISR2003, Portokalidis:GlobalISR2011}, address space layouts~\cite{Shacham:ASR2004}, IP addresses~\cite{Al-Shaer13, antonatos2007defending, Jafarian:OFRHM2012, Kewley:DyNAT2001, Sharma18}, port numbers~\cite{Luo:RPAH2015}, proxies~\cite{jia2013motag}, virtual machines~\cite{zhang2012incentive, ben2016attacker}, operating systems~\cite{Thompson:OSRotation2014}, or software programs~\cite{Jackson:Compiler2011}. Table~\ref{table-movingelements} summarizes the moving elements by MTD techniques in different system layers~\cite {alavizadeh2017effective, hong2016assessing}. A change in the system configuration attributes must cause a change on the attack surface of the system leading to the loss of work and increased complexity for attackers. Additionally, the number of possible values the dynamic attribute can change to must be vast, thwarting the attacker's ability to simply brute-force the future value~\cite{Cai16}.

\item {\bf How to move}: `How to move' defines how to change the moving attributes (i.e., targets) to increase unpredictability and/or uncertainty, leading to an attacker's high confusion. This is related to the three operation-based MTD classification components: shuffling, diversity, and redundancy (SDR). Common MTD techniques include artificial diversity~\cite{ben2016attacker, huang2011introducing, Jackson:Compiler2011, Rowe:Aartificial2012, Van:Noncespaces2009} or randomizations~\cite{Al-Shaer13, antonatos2007defending, Jafarian:OFRHM2012, jia2013motag, Kc:ISR2003, Kewley:DyNAT2001, Luo:RPAH2015, MacFarland15, Shacham:ASR2004, Sharma18, Skowyra:MTD'16, Wang:MTD'17}, which rearrange or randomize the various system and network attributes where each work can belong to shuffling, diversity, and/or redundancy techniques. Fig.~\ref{fig:how_to_move} depicts examples of how to move the moving elements using MTD based on the SDR classification~\cite {alavizadeh2017effective, hong2016assessing}.

\item {\bf When to move}: `When to move' involves deciding the optimal time to change from the current state of an MTD system to a new state, invalidating information or progress gained by an attacker in the current state. The \new{three common} 
types of adaptations are reactive, proactive, or hybrid strategies. The {\em reactive adaptation} is based on an event or an alert
that executes the adaptation as a response to a message from the detection of suspicious activity.  Reactive MTD attempts to counter actions taken by an attacker under the assumption that the attacker's actions are detected. The {\em proactive adaptation} triggers MTD techniques by changing configurations or properties 
on a schedule, with either fixed or random intervals between adaptations. Proactive adaptations ensure that any information gained by the attacker quickly expires \cite{MacFarland15}, thus regular movement is important. Proactive MTD moves regardless of the presence of an attacker costing additional delay in the protected system. The {\em hybrid adaptation} is based on both reactive and proactive features, wherein the time interval to trigger an MTD operation is adaptive upon certain events or security alerts while the interval is also bounded in length to prevent potential, undetected security threats~\cite{zhuang2013investigating}.  Fig.~\ref{fig:when_to_move} shows an example of `when to move' which addresses at what time point the element(s) of MTD can be moved under hybrid adaptations~\cite {alavizadeh2017effective, hong2016assessing}. In Fig.~\ref{fig:when_to_move}, a blue line means an MTD can be triggered at a fixed MTD interval whereas an orange line shows an MTD triggered by an event (e.g., an intrusion detection alert). Intuitively, we can notice that using the fixed MTD interval is more proactive than using the adaptive MTD interval in confusing attackers. However, high frequency of MTD operations may incur high cost. Hence, an optimal MTD interval can be identified at run-time to balance both cost and security. A running system can have different moving elements (e.g., IP, Port, OS, VM, Applications) over the time, and their changes can take place either at a regular fixed interval of time or upon receiving an intrusion detection alert.
\end{itemize}

\subsection{Discussions: Benefits and Caveats of MTD} 

In this section, we discuss the key benefits of MTD and caveats in developing MTD techniques, in relation with the roles and design principles of MTD techniques.

The {\bf key benefits} of MTD techniques include as follows:
\begin{itemize}
    \item {\bf Providing affordable defense opportunities.} Compared to 
    conventional security mechanisms aiming to perfectly eliminate any vulnerabilities and risks that can be introduced by attackers, MTD provides a new perspective of a defense system by continuously changing the attack surface, which makes 
    it harder for attackers
    to achieve their goals. MTD allows us to leverage legacy system components and existing 
    technologies, enabling a greater likelihood of achieving affordable defense solutions and avoiding the necessity of
    creating a new, highly robust security mechanism, such as cryptographic solutions, which 
    requires more resource / time to develop but may be less applicable in broad contexts (e.g., under the assumption of trusted entities for key management). 
    \item {\bf Providing another layer of defense in cooperation with existing defense mechanisms.}
    At the same time, MTD can cooperate with other defense mechanisms by assisting in intrusion detection that can thwart potential attackers and/or provide new attack patterns observed during the reconnaissance stage of the attack. In addition, MTD can add another layer of defense when a defensive deception is used to deceive an attacker. In particular, when the attacker realized the deception and can successfully identify vulnerable system component, not being lured by the deception (e.g., a honeypot), MTD can be executed to migrate the system platform in order to invalidate the information of the system vulnerabilities collected by the attacker in the current system configuration.
\end{itemize}
However, we also need to be aware of the following {\bf caveats in developing MTD techniques}:
\begin{itemize}
    \item {\bf High frequency of triggering MTD operation may significantly hinder seamless, quality service provision to users.} This issue is related to a design principle related to `when to move' a target because triggering MTD operations too often naturally leads to reducing service availability and/or interrupting seamless service provision.
    \item {\bf Non-adaptive, non-intelligent MTD may waste defense cost while not decreasing system vulnerabilities.} In some system / network settings, resources are highly constrained, such as IoT environments or mobile, wireless 
    networks. The strategy in deciding how to expend defense resources is critical to prolonging system lifetime and increase system reliability (or survivability). If a deployed MTD is not adaptively and/or intelligently executed based on a level of detected security vulnerabilities, then the required system goals, such as minimizing defense cost and maintaining a certain level of system security, may not be met.
    \item {\bf How to execute MTD operations in resource-constrained, distributed network environments has not been sufficiently studied in the literature.} Most existing MTD approaches consider a trusted third party for network management associated with running an MTD technique. However, in some decentralized or distributed, resource-constrained environments, such as mobile ad hoc networks, wireless sensor networks, or IoT networks, it is not easy to identify a trusted infrastructure that can make critical decisions on MTD executions. Further, the cost of executing MTD operations should be considered as a priority as those contested environments can only afford highly lightweight MTD solutions.
\end{itemize}

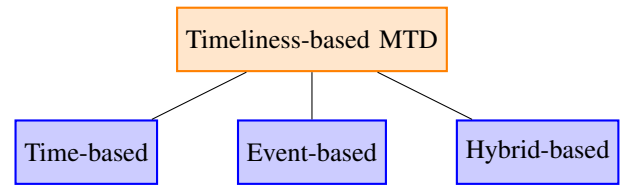
\begin{figure}[t!]
	\centering
\begin{tikzpicture}
\tikzstyle{every entity}=[fill=blue!20,draw=blue,thick]
\tikzstyle{every relationship}=[fill=orange!20,draw=orange,thick,aspect=1.5]
\node[entity] (time) at (0,0) {Time-based};
\node[entity] (event) at (3,0) {Event-based};
\node[entity] (hybrid) at (6,0) {Hybrid-based};
\node[entity, fill=orange!20,draw=orange,thick] (timeliness) at (3, 1.5) {Timeliness-based MTD};
\draw (timeliness) -- (time);
\draw (timeliness) -- (event);
\draw (timeliness) -- (hybrid);
\end{tikzpicture}
	\caption{Classifications of Timeliness-based MTD.}
	\label{fig:timeliness-based-MTD}
\end{figure}

\section{Classification Types of MTD} \label{sec:mtd_classification}
MTD techniques have been studied under various classifications with different criteria. In this section, we discuss how the MTD techniques have been classified in the literature. In addition, we distinguish the concepts of MTD from those of deception and discuss what the commonalities and differences are between them.

\subsection{Timeliness-based MTD} \label{subsubsec:time-mtd-classification}
Timeliness-based MTD classifies MTD techniques based on criteria to determine `when to move.' Fig.~\ref{fig:timeliness-based-MTD} depicts the three common timeliness-based MTD categories as follows:
\begin{itemize}
\item \new{\textbf{Time-based}: This approach triggers an MTD operation based on a certain time interval called the {\em MTD interval} which can be a fixed interval or a variable interval~\cite{Cai16}.  With a fixed time interval, the MTD mechanism periodically changes the attack surface (e.g., IP/Port addresses, OS rotation, VM migration) with a constant equal time which remains unchanged. If the interval time is too large, then an attacker may be allowed a sufficient amount of time to scan a system and then penetrate into the system, resulting in a security breach. On the other hand, if the interval time is too small, then the MTD triggers even when there is no attack on the system, wasting defense resources and degrading performance. 
Therefore, determining the MTD interval time to perform an MTD operation has a significant impact on the effectiveness and efficiency of a given MTD technique~\cite{Carroll14}.}
\item {\bf Event-based}: This approach performs an MTD operation only when a certain event occurs. The event can encompass any indication an attacker accesses a system or attempts to launch a certain attack. That is, if the defender can accurately predict potential attacks, then events can trigger the appropriate MTD operation. In order to discover the key events that should trigger the MTD action, attack prediction based adaptive MTD approaches have been proposed using machine learning~\cite{colbaugh12}, game theory~\cite{zhu2013game}, and control theory~\cite{Rowe:Aartificial2012}. 
\item {\bf Hybrid}: Some MTD approaches take a hybrid strategy. These approaches ~\cite{ Keromytis2012MEERKATS, Rowe:Aartificial2012, zhuang2012simulation, zhuang2013investigating} execute MTD operations adaptively based on both the time and event-based MTD strategies for proactive and reactive adaptations, respectively. 
\end{itemize}

\subsection{Operation-based MTD} \label{subsec:goal-mtd-classification}
 Operation-based MTD classifies MTD techniques based on criteria to determine `how to move'.
 \citet{hong2016assessing} 
 labeled three types of MTD techniques based on the nature of the operations: {\em shuffling}, {\em diversity}, and {\em redundancy}. Each type is detailed as follows:
\begin{itemize}
\item {\em Shuffling}: This technique rearranges or randomizes system configurations, such as mutating IP addresses at a TCP/IP layer or dynamically adjusting the migration time of VMs. The key goal of these shuffling-based MTD techniques is to increase confusion and uncertainty for attackers 
(i.e., to make the identification of vulnerable targets more difficult) by making information collected by the attackers obsolete or by wasting attackers resources in the collection of useless information.  Ultimately, shuffling-based MTD can prevent or delay the attackers from accessing a target system. 
Because the system earns more time to monitor the attack behaviors (e.g., scanning attacks), the system's defense mechanisms (e.g., IDS) can prepare more intelligent strategies to deal with the attack based on identified attacks. 
\item {\em Diversity}: This technique employs the deployment of system components with different implementations that provide the same functionalities. The examples include the use of diverse paths for routing or the change of platforms consisting of different implementation of software components or migration between different platforms (i.e, software stacks and/or hardware).  Diversity-based MTD aims to enhance system resilience by increasing fault-tolerance in that the system can provide normal services in the presence of attackers in the system. 
\item {\em Redundancy}: This technique provides multiple replicas of system (or network) components, such as multiple paths between nodes in a network layer or multiple software components providing the same functionality at the application layer. The key aim of redundancy-based MTD is to increase system dependability (e.g., reliability or availability) by providing redundant ways of providing the same services when some of the network nodes or system components are compromised. In this sense, 
redundancy contributes to increasing system (or network) resilience in \new{the presence of insider threats}. This technique can often be combined with diversity-based MTD so that, for example, redundant services are available where the attackers are required to know additional credentials or intelligence to use other alternative components (e.g., different or additional credentials or a different level of privileges). 
\end{itemize}

\begin{figure}[t!]
	\centering
\begin{tikzpicture}

\node[shape=ellipse,draw=orange, align=center, fill=orange!20]  (S) at (-2.5, -2) {Shuffling};

\node[shape=ellipse,draw=green, align=center, fill=green!20]  (D) at (0, 0) {Diversity};

\node[shape=ellipse,draw=blue, align=center, fill=blue!20]  (R) at (2.5, -2) {Redundancy};

\node[shape=rectangle,draw=brown, align=center, fill=brown!50]  (SG) at (-2.5, -3.5) {Performance / Efficiency};

\node[shape=rectangle,draw=green, align=center, fill=green!50]  (DG) at (0, 1) {Resilience / Robustness};

\node[shape=rectangle,draw=blue, align=center, fill=blue!50]  (RG) at (2.5, -3.5) {Reliability / Availability};

\draw (D)--(DG);
\draw (S)--(SG);
\draw (R)--(RG);

\draw[>=triangle 45, <->, black, thick] (D) to [bend right=30] (S);

\draw[>=triangle 45, <->, black, thick] (D) to [bend right=-30] (R);

\draw[>=triangle 45, <->, black, thick] (R) to [bend right=-30] (S);

\end{tikzpicture}

	\caption{Relationships between Shuffling, Diversity and Redundancy (SDR).}
	\label{fig:sdr}
\end{figure}
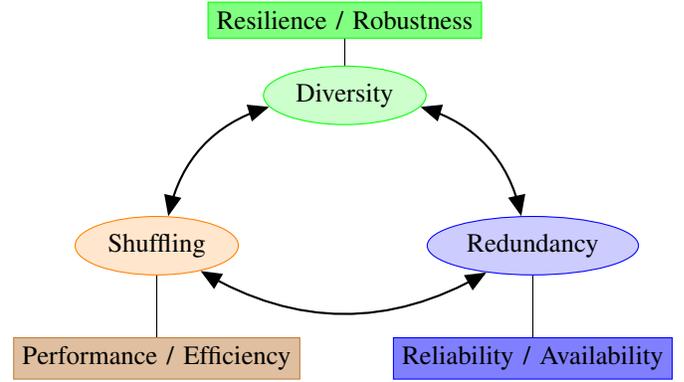

Although these three types of MTD operations can be used to classify MTD techniques (see Section \ref{sec:mtd_techniques}), they often support each other directly or indirectly in realizing their respective goals. We summarize the connections between the SDR types and their goals in Fig.~\ref{fig:sdr}.

First, the concept of {\em diversity} is adopted from the principle that the diversity of system components can enhance security (i.e., software polyculture enhances security~\cite{franz2010unibus, gherbi2011diversity, taylor2005diversity}), leading to high system resilience (or robustness) even in the presence of attacks. If system diversity is high, there exists a variety of alternatives to the provision of the same service, such as multiple routes from a source to a destination to deliver a message. 

Second, {\em shuffling} also can enhance diversity because how to rearrange system components can affect the degree of system diversity. For example, a platform migration using a shuffling technique leads to higher diversity. How to shuffle targets is closely related to how much cost incurs from shuffling the targets and, accordingly, how much it can enhance security or interrupt services to normal users. The effectiveness of shuffling is also affected by how diverse a target is in nature. For example, given a network, we want to increase the number of different types of software that provides the same set of services. But if we only have one or two software types, then there is an inherent limitation of the performance achievable by the shuffling operation. On the other hand, if the shuffling is optimally performed, system diversity also increases, leading to higher system resilience against attacks.

Finally, high diversity and effective shuffling are critical to increasing redundancy for achieving higher reliability (or availability) while properly dealing with attackers. Particularly, if redundancy is high with low diversity due to a poor or no shuffling technique, it can be also leveraged by the attackers to persistently use alternate ways to get into the system. Therefore, these three types of techniques, shuffling, diversity, and redundancy, namely SDR, impact each other and should be considered in combination to properly enhance system security or resilience.

\subsection{MTD vs. Deception} \label{subsec:deception_mtd}

MTD and deception techniques have been used without clear distinction. To clarify this distinction, we briefly describe what deception is and how it has been used as a defensive technique in the literature. More importantly, we discuss what are different and common between MTD and defensive deception in terms of their purposes and strategies.

\subsubsection{Defensive Deception}
In military settings, deception is originally derived from actions to be executed to intentionally mislead an opponent's decision associated with strengths and weakness of military capacities, intents, and/or operations. Defensive deception is created to affect an attacker's action in a way a defender wants to achieve its mission~\cite{sharp2006MilitaryDeception}. Deception has been studied based on a number of classifications.

\citet{bell1991cheating} and \citet{almeshekah2016cyber} discuss deception techniques in terms of either `hiding the real' or `showing the false.' That is, the defender can modify information and/or present misleading information to the attacker. For example, the defender can mislead the attackers to misjudge by providing a false sense that the attacker has complete, certain, relevant information for its decision making, which is not true indeed.

\citet{caddell2004deception} classifies deception in terms of `passive deception' and `active deception.' The passive deception includes actions to hide capabilities or intents from the attacker. The active deception involves leading the attacker to form false beliefs by providing false information. \citet{daniel1982strategic} introduce two types of deception techniques based on their goals, aiming to increase an attacker's perceived ambiguity and to mislead the attacker to misjudge.

\new{\citet{rahman2013game} proposed a game-theoretic approach using Nash equilibrium strategy as a deception technique to prevent remote OS fingerprinting attacks. They showed that the designed technique named \emph{DeceiveGame} can significantly decrease the fingerprinter attack success probability while the overall usability of the system is preserved without performance degradation.} \citet{bell1991cheating} and \citet{almeshekah2016cyber} show some other example deception techniques that dissimulate information in order to hide real information and simulate information to make the attacker form false beliefs. Examples of dissimulation deception are masking (i.e., hiding the real), repackaging (i.e., creating the false), and dazzling (i.e., confusing an attacker)~\cite{almeshekah2016cyber, bell1991cheating}.
Examples of the simulation deception include mimicking (i.e., imitating the real), inventing (i.e., creating the false), and decoying (i.e., luring)~\cite{almeshekah2016cyber, bell1991cheating}.
Based on our discussion above, we can find the commonalities and differences between deception and MTD as below.


\subsubsection{Commonalities between MTD and Deception} 
Both approaches have the same goal to confuse attackers by increasing their uncertainty in the decision process of launching attacks. Hence, the deception techniques used for increasing confusion can also belong to MTD. In addition, even if the deception techniques are used to mislead attackers but they are also being dynamically deployed (i.e., a different set of decoy nodes is deployed at a different time~\cite{Ge18}), they can be treated as MTD techniques.

\subsubsection{Differences between MTD and Deception} 
Although these two defense techniques are common in their goal, deception takes a more aggressive strategy than MTD in terms of intentionally presenting false information for the attacker to form false beliefs while MTD does not use a strategy of lying or disseminating false information to mislead attackers. But these two defense mechanisms can work together well to benefit each other in the following sense. Deploying deception can be less costly than MTD operations and can help MTD to save its cost because the attack can be delayed if an attacker is successfully deceived by the deployed defensive deception and accordingly MTD does not have to be triggered as often as the system without deception~\cite{Cho18}. In addition, the defense strength achievable by defensive deception is limited in nature because the deception can be ultimately known to the attacker sometime later. When the attacker found it was deceived by the defensive deception, the system can trigger an MTD operation to protect the system without security breach.

\subsection{Discussions on MTD Classifications} 

While the timeliness-based MTD categorizes MTD techniques based on the principle of `when to move,' the operation-based MTD classifies MTD techniques by answering `how to move.' In the timeliness-based MTD, when a defender has high uncertainty about an attacker's current activities and/or behaviors, it is better to trigger an MTD operation based on a certain, fixed time interval. However, after the defender becomes certain about the attack patterns / activities, it can detect the system's vulnerabilities more accurately. And then, the defender can adaptively execute the MTD operation to save defense cost by using an event-based MTD, for example, triggering an MTD operation based on intrusion alerts.

In the operation-based MTD classification, a single SDR type can be used in an MTD technique or multiple types can be used as we discussed in Section~\ref{subsec:mtd_tech_hybrid}. Although a hybrid approach may achieve more enhanced system security, it may introduce additional defense cost and/or issues related to service availability. Again, identifying a critical tradeoff point between multiple, conflicting system goals is a non-trivial task in developing a cost-effective and efficient MTD solution.

These two classification methods answer two different design questions in developing MTD techniques (i.e., \new{how}
-to-move and when-to-move). Hence, we can even consider both operation-based and timeliness-based classification in order to develop a secure, affordable, adaptive, and QoS-aware MTD solutions.

\section{MTD Techniques based on Shuffling, Diversity, and Redundancy} \label{sec:mtd_techniques}
In this section, we survey MTD techniques based on the operation-based MTD classification with the three categories, including shuffling, diversity, and redundancy. In addition, since some approaches take hybrid approaches combining more than one MTD technique, we categorize these as `hybrid' in this paper. When some deception techniques are used to change the attack surface, we also discuss them as MTD techniques based on the used classification criteria.

\subsection{Shuffling} \label{subsec:mtd_tech_suffling}
\new{A shuffling technique rearranges or randomizes system configurations. In this section, we discuss various ways to change the system or network configurations.}
\subsubsection{IP Shuffling / Mutations / Host Randomization} Many shuffling-based MTD techniques used IP shuffling or mutation in various network domains. \citet{Sharma18} developed IP shuffling MTD by using the concept of IP multiplexing (or demultiplexing) in an SDN environment. \citet{MacFarland15} used a host IP address mutation to defend a large-scale network by employing an SDN controller that controls DNS interactions. \citet{antonatos2007defending} proposed an IP randomization technique to thwart hitlist worms attacks, aiming to avoid malicious worms that gather information about victim targets in a networked system and making it hard for attackers to identify new, vulnerable targets.  \citet{Jafarian:OFRHM2012} implemented an IP shuffling technique mutating the IP addresses unpredictably while minimizing the overhead of the MTD operations. The authors used the SDN-based OpenFlow (OF) protocol that frequently assigns Virtual IP addresses translated from and/or to a real host's IP address. 

\citet{Carroll14} also studied network address shuffling to protect honeypots as an MTD technique based on probability models to quantify the attack success probability with respect to the size of a network, the number of addresses scanned by attackers, the number of system vulnerabilities, and the frequency of triggering the shuffling-based MTD.  \citet{kampanakis2014sdn} employed host randomization and mutation, evaluating the performance in terms of the attackers' overhead increased by the MTD technique used in an SDN platform. 

\subsubsection{Port Hopping}
\citet{luo2014effectiveness} studied a port hopping technique based on the concept of shuffling-based MTD to effectively deal with reconnaissance attacks by hiding service identities while increasing confusion for potential attackers. This work quantifies the effectiveness of the proposed port hopping technique based on the attack success probability 
as a function of the values of key design parameters, including the size of a port pool, the number of probes, the number of vulnerable services, and the frequency of port hopping. 

\subsubsection{Packet Header Randomization} \citet{Wang:MTD'17} proposed a technique called U-TRI, Unlinkability Through Random Identifier, which uses the randomization of a packet header identifier to confuse attackers.

\subsubsection{Network Topology Shuffling} The underlying idea of this technique is to invalidate an attacker's path information by continuously changing routes in networks. 
\citet{Achleitner:2017Deceiving,achleitner2016cyber} 
proposed a virtual topology generation framework against network scanning attacks by leveraging the SDN technology.  \citet{Hong:2016Optimal} presented an optimal network reconfiguration technique based on the concept of shuffling-based MTD for SDN environments. They solved a shuffling assignment problem and showed the increase in network security when network routing paths are diversified. 
\subsubsection{VM / Proxy Migration}	
\citet{danev2011enabling} considered 
securely migrating Virtual Machines (VMs) as an MTD mechanism in a private cloud. The underlying idea of this work is to utilize an extra physical Trusted Platform Module (TPM) and trusted parties for the migration process as well as public key infrastructure to secure the protocol. \citet{zhang2012incentive} developed a periodic VM migration as an MTD technique based on the balance between the level of security obtained and the cost incurred upon the migration of VMs. 

\citet{penner2017combating} developed a set of MTD technologies to change the location of VMs in a cloud to defend against attacks leveraging a Multi-Armed Bandit (MAB) policy, which aims to exploit VMs providing the highest reward. The proposed MTD techniques are deployed to minimize the MAB attacks launched by the attacker aiming to obtain credentials (e.g., credit card information) or critical data. The proposed MTD techniques are evaluated based on how much time it takes to switch VMs. However, the enhanced security introduced by the MTD techniques has not been investigated.  \citet{jia2013motag} devised an MTD technique to deal with Distributed Denial-of-Service (DDoS) attacks through securing data transmission between authenticated (legitimate) clients and a protected server. This method used the technique called {\em client-to-proxy shuffling} that continuously moves secret proxies. For evaluation, they assessed the resistance of their method toward brute-force attacks, as well as proxy-based and communication-based overhead.

\citet{peng2014moving} proposed VM migration or snapshotting and diversity or compatibility of migration for cloud networks. This work considered attackers' learning based on accumulated intelligence obtained and dynamics and heterogeneity of a service's attack surface. Finally, this work proposed a probability model to propose an MTD service deployment strategy. This study found that the proposed MTD technique is more effective under a dense service deployment and strong attacks while the defender's awareness toward the heterogeneity and dynamics of the attack surface is helpful for enhancing the effectiveness of the MTD as it can be used to determine when/how to perform the MTD.  

\subsubsection{Software / Service Reconfiguration} 
\citet{Casola2013IRI} developed an MTD technique based on the reconfiguration of devices in an IoT environment whose cryptosystems and firmware versions are shuffled. \citet{vikram2013nomad} proposed a shuffling based MTD technique on the application layer for a secure web through randomizing HTML elements. Most bots attacking the web use static HTML elements in the HTTP content/form page. Hence, the randomization of those HTML elements and parameters could be an appropriate technique for avoiding web bot attacks.
\new{
\subsubsection{Platform Diversity}
\citet{okhravi2012creating} proposed the Trusted Dynamic Logical Heterogeneity System (TALENT) framework for live-migrating of the critical infrastructure applications across heterogeneous platforms, which permits running of the critical applications to change its hardware and operating system (OS). It changes a platform on-the-fly, creates a cyber moving target, and provides the cyber survivability through platform diversity. TALENT creates a virtualized environment at the OS-level using containers with a checkpoint compiler for migrating a running application of different platforms. It preserves the state of the application, such as the execution state, the open files, and the network connections during the migration.
}
\new{
\subsubsection{OS-Rotation:} \citet{Thompson:OSRotation2014} developed the Multiple OS Rotational Environment (MORE) MTD based on the existing technology to achieve a feasible MTD solution at an OS-level. MORE MTD consists of a set of virtual machines (VMs) equipped with a different distribution of OS (i.e., different Linux distributions) and web applications.  The periodic rotation of the hosts/VMs creates the dynamic environment controlled by an administrator's machine running a daemon process.  This technique reduces the likelihood of a successful exploitation of the OS's vulnerabilities and its security impact by limiting the duration of the rotation window. The ``rotation window'' is the duration of an OS being exposed and vulnerable to an attack.
}

\vspace{1mm}
\noindent {\bf Pros and Cons:} Shuffling-based MTD techniques can easily work with existing technologies without developing another security mechanism that 
requires a thorough security analysis. Hence, in terms of the development cost and resources, 
these methods are useful, immediately applicable, and highly compatible with legacy devices and technologies.
However, since shuffling relies on the quality of existing technologies, if those existing technologies are not robust enough against attacks (i.e., well-known vulnerabilities to attackers), the security achieved by this shuffling technique can be significantly limited by those vulnerabilities of the legacy devices/technologies.
In this case, the shuffling frequency can be increased to help cover those vulnerabilities, but this also will increase the defense cost incurred by over-utilizing MTD as another layer of defense. Furthermore, the shuffling spaces (e.g., a number of virtual IPs that can be assigned for a real IP) are critical to enhancing security.

\subsection{Diversity} \label{subsec:mtd_tech_diversity}
\new{Diversity-based MTD techniques employ the deployment of system components with different implementations. In this section, we discuss how system diversity is realized based on a variety of domains.}

\subsubsection{Software Stack Diversity}
\citet{huang2011introducing} and \citet{huang2010security} introduced a diversity MTD technique in order to enhance network resilience and service provisions. They deployed a diversity method to the virtual servers (VS) like OSs, virtualization components, web servers, and application software. They evaluated the proposed diversity method with respect to attack success probability. 

\subsubsection{Code Diversity}
\citet{azab2011chameleonsoft} developed an MTD technique that changes a running program's variants erratically based on the concept of diversity. The proposed method divides a large program into components (e.g., cells or tasks) that can be performed by functionally-equivalent variants. Their method includes a recovery mechanism to enhance the resilience of the proposed technique. Choosing a different variant at runtime makes 
it hard for an attacker to penetrate into and scan the system. The proposed method can mitigate even the effect of the attack success by affecting one variant that can be instantly replaced by another variant through the recovery mechanism. However, the robustness against attacks based on security analysis is not clear while its implementation is quite complex which can be questioned for applicability in diverse contexts. 

\new{In automated software diversity, a program can diversify at instruction, block, loop, program and system levels in different phases of the life cycle, such as compilation and linking, or installation~\cite{Larsen:SoK2014}.  A JIT (just-in-time) compiler (i.e., Java/Python JIT compilers) generates code during the execution of the program. The JIT compilers can be used for the code diversification. In this case, the security of the system depends on how predictable the  JIT compilers are.  The code diversity can be enhanced with adaptive notification-oriented programming (i.e., assembly language instruction, no-op) ~\cite{Jangda:JITCodeDiversification2015}.}
\subsubsection{Programming Language Diversity}
\citet{taguinod2015toward} focused on an MTD technique based on the diversification of programming languages to avoid code and SQL injection attacks. This work proposed the MTD technique to be applied in different layers of the web application to change the implemented language of the web application without affecting or disrupting the system functionality.
	
\subsubsection{Network Diversity}
\citet{zhuang2012simulation} investigated the relationship between the diversity of network configurations and attack success probability by evaluating the MTD technique based on a logical mission model. The authors examined how the network size, the frequency of shuffling / adaptations, and the number of vulnerable computers affect the effectiveness of the MTD technique based on the experimental results tested from the network security simulator, {\tt NeSSi2}.

\vspace{1mm}
\noindent {\bf Pros and Cons:} Similar to shuffling-based MTD, diversity-based MTD can leverage the existing technologies. As discussed earlier, diversity-based MTD techniques are often combined with shuffling-based counterparts to double the effectiveness of a proposed MTD. However, since the diversity-based MTD is also based on the existing technologies, if those technologies have high vulnerabilities to attacks, introducing MTD as another layer of defense does not provide better security, even with the sacrifice of additional defense cost. In addition, the degree of diversity significantly affects the effectiveness of the MTD. For example, if there are not many alternatives to use (e.g., only two versions of software that provides the same functionality with different implementations), there will be inherent limitations in enhancing the security.

\begin{table*}[t!]
    \caption{A Summary of Pros and Cons of MTD Techniques}
    \label{tab:mtd-techniques}
    \vspace{-2mm}
    \centering
    \begin{tabular}{|P{1.5cm}|P{4.5cm}|P{4cm}|P{4cm}|P{2.5cm}|}
    \hline
         {\bf Type} & {\bf Main Techniques} & {\bf Pros} & {\bf Cons} & {\bf Ref.}\\ \hline
         Shuffling & IP shuffling, mutations; host randomization packet header randomization; network topology shuffling; VM / proxy migration; software / service reconfiguration 
         & Leveraging legacy devices or technologies; providing affordable, economical defense
         & Potential high cost or service interruptions if not executed adaptively; limited by the inherent vulnerabilities of the existing technologies used or shuffling spaces
         & \cite{Achleitner:2017Deceiving, achleitner2016cyber, antonatos2007defending, Carroll14, Casola2013IRI, danev2011enabling, Hong:2016Optimal, Jafarian:OFRHM2012, jia2013motag, kampanakis2014sdn, luo2014effectiveness, MacFarland15,  peng2014moving, penner2017combating, Sharma18, vikram2013nomad, Wang:MTD'17, zhang2012incentive}
         \\ \hline
         Diversity & Software stack diversity; code diversity; programming language diversity; network diversity 
         & Leveraging legacy devices or technologies; working well with shuffling-based MTD to double the effect
         & Limited by the inherent vulnerabilities of the existing technologies or a number of software alternatives
         & \cite{azab2011chameleonsoft, huang2011introducing, huang2010security, taguinod2015toward, zhuang2012simulation}
         \\ \hline
         Redundancy & Redundancy of software components; redundancy of network sessions & High reliability/availability achievable; easily combined with shuffling or diversity to significantly secure a system and ensure service availability
         & Additional cost to move / setup additional system components (e.g., servers or routing paths); if not properly executed, it increases attack surface.
         & \cite{Li14, yuan2013architecture}
         \\ \hline
         Hybrid 
         &  Redundancy and diversity based MTD to provide diverse replicas of web services; shuffling and redundancy for VM migration with multiple VM replicas; combining shuffling, redundancy, and diversity based MTD for cloud-based web services
         & Providing enhanced security by combining multiple MTD techniques which a single solution cannot introduce 
         & Additional cost; high complexity for multi-objective optimization with multiple operational decisions
         & \cite{alavizadeh2017effective, Alavizadeh18, alavizadeh2018evaluation,  gorbenko2009using}
         \\ \hline
    \end{tabular}
\end{table*}
	
\subsection{Redundancy} \label{subsec:mtd_tech_redundancy}
\new{Redundancy techniques provide multiple replicas of system or network components providing the same functionality at the network/application layer. In this section, we discuss how these techniques are studied in the existing approaches.}

\subsubsection{Redundancy of Software Components}
\citet{yuan2013architecture} proposed a redundancy method for web servers to prevent malicious code injection attacks on a web server by developing a self-protection model, including architectural adaptation threat detection and mitigation. They used the so-called {\em agreement-based redundancy} that provides replicas of software components at runtime. However, this work did not investigate the effectiveness of the proposed MTD mechanism.

\subsubsection{Redundancy of Network Sessions}
\citet{Li14} proposed a traffic morphing mechanism by adopting the concept of MTD in terms of redundancy for cyber-physical system (CPS) environments. The proposed traffic morphing algorithm is designed to protect CPS sessions by maintaining a number of redundant network sessions which have the distributions of inter-packet delays indistinguishable from those observed in normal network sessions. A CPS message will be disseminated via one of these sessions to meet its given time constraint. In the process of dynamically adjusting the morphing process, this work showed the low complexity of the proposed work and high adaptivity to the dynamics of CPS sessions.

\vspace{1mm}
\noindent {\bf Pros and Cons:} Unlike shuffling or diversity based MTD, redundancy-based MTD has a greater bearing on service availability to users, which is often measured by some concept of system dependability (e.g., availability or reliability) in which the availability (or reliability) can be easily interrupted by performing shuffling or diversity based MTD, particularly incurring interrupted services. Note that redundancy-based MTD can be well-mingled with the other two techniques with the aim of increasing both security and performance. However, creating additional replicas of system components (e.g., servers or paths) incurs an extra cost. In addition, if redundancy-based MTD is not properly executed, it provides an even greater chance for an attacker to perform attacks on a larger attack surface (i.e., another server to attack or another path to reach a target) than the system without the redundancy-based MTD.   

\subsection{Hybrid} \label{subsec:mtd_tech_hybrid}

Hybrid MTD combines multiple MTD techniques among the types (i.e., shuffling, diversity, and redundancy) to work in cooperation.  In this section, we discuss how 
MTD techniques can be integrated to enhance system security and performance.

\subsubsection{Diversity + Redundancy (D+R)}
\citet{gorbenko2009using} developed a web service based on the concept of redundancy and diversity by providing diverse replicas of web services for maximizing system dependability, which is assessed by availability, reliability, and/or service response time. However, this work does not conduct an in-depth investigation of security analysis.

\subsubsection{Shuffling + Redundancy (S+R)}
\citet{alavizadeh2017effective} combined shuffling with redundancy by developing shuffling-based VM Live-Migration (VM-LM) and generating VM replicas to enhance both security and dependability of a system. The authors assessed the performance of the developed MTD technique based on the metrics of System Risk ($Risk$) and Reliability ($R$) to show how effective the combined MTD technique is in a cloud environment. 
Similarly, \citet{alavizadeh2018evaluation} conducted a substantial analysis to evaluate the effectiveness of two other combinations of MTD techniques through considering three key metrics, including Risk ($Risk$), Attack Cost ($AC$), and Return on Attack ($RoA$). Based on their definition, an appropriate MTD technique should decrease $Risk$ and $RoA$  while increasing $AC$. The authors showed that combining shuffling and diversity can optimally meet these multiple objectives, whereas a single solution with either shuffling or diversity cannot. 

\subsubsection{Shuffling + Diversity + Redundancy (S+D+R)}
\citet{Alavizadeh18} combined the three MTD techniques (i.e., S+D+R) in the virtualization layer of a cloud, in order to quantify the cloud's security level through comprehensive security analysis. Similar to \cite{alavizadeh2018evaluation}, the authors investigated the performance of the combined MTD techniques based on formal graphical security models for the cloud environment using the three metrics (i.e., $Risk$, $RoA$, and $AC$) plus system availability ($SA$).

\vspace{1mm}
\noindent {\bf Pros and Cons:} Hybrid MTD approaches based on more than one MTD technique can introduce additional benefits of enhancing security, which may not possible under a single technique based MTD solution. In particular, diversity or redundancy combined with shuffling can significantly increase security while decreasing defense cost or service interruptions. For example, MTD combining shuffling with diversity may not require shuffling as frequent as an MTD with a shuffling only solution because the diversity of system components can make it harder for attackers to figure out their vulnerabilities.  In addition, redundancy can increase system availability leading to high service quality provided to users with fewer interruptions. However, as discussed earlier, it may introduce a larger attack surface than a single MTD. It also requires additional overhead and/or complexity in combining multiple techniques as a single solution and encounters a complex, multi-objective optimization problem with multiple operational constraints, which is known to be an NP-hard problem~\cite{Cho17-moo}.

\subsection{Discussions on Operation-based MTD Techniques} 
From our literature review on MTD techniques based on the three MTD classes (shuffling, diversity, and redundancy), we observed the following:
\begin{itemize}
    \item {\bf A large volume of shuffling-based MTD techniques}: Many shuffling-based MTD techniques, such as platform migration or network topology changing, are combined with diversity-based techniques so that their positive effect is maximized to enhance security. However, critical tradeoffs between security (e.g., security vulnerabilities reduced) and performance (e.g., defense cost or service availability) have not been thoroughly investigated.
    \item {\bf The improvement of security bounded by the vulnerabilities of existing technologies leveraged by MTD}: Due to the nature of MTD leveraging existing technologies, there is an inherent limitation in achieving security if the existing technologies do not have a sufficient level of security robustness. For example, the vulnerabilities in the leveraged technologies (e.g., software vulnerabilities) cannot be removed even if a diverse software stack is used.
    \item {\bf Redundancy-based MTD combined with other techniques to ensure service availability}: We noticed that the redundancy-based MTD does not play a significant role as proactive MTD defense compared to the other two techniques. Rather, redundancy-based MTD is used to enhance service availability (or reliability), which can be easily excluded
    when shuffling/diversity techniques are solely used without consideration of the critical tradeoff between enhanced security and defense cost introduced.
    \item {\bf Highly promising hybrid MTD with further investigation to optimize its deployment}: Hybrid MTD approaches are very promising because they can achieve multiple objectives of maximizing security with minimum defense cost or maximum service availability when they are properly implemented. However, additional overhead or complexity introduced by combining multiple techniques is not avoidable and it is necessary to conduct a very careful examination about a hybrid MTD before it is deployed.
\end{itemize}

We summarize the key techniques under each class (i.e., shuffling, diversity, redundancy, or hybrid) of MTD techniques and their pros and cons in Table~\ref{tab:mtd-techniques}.

\section{Key Attacks Mitigated by MTD and Limitations of Attack Modeling in MTD} \label{sec:threat-models}
In this section, we discuss the characteristics of attacks considered in the state-of-the-art MTD techniques. To be specific, we discuss: (1) the characteristics of advanced attacks; (2) the cyber kill chain model; (3) attack types mainly considered in the existing MTD approaches; and (4) limitations of the existing threat models.

\subsection{Characteristics of Advanced Attacks} \label{subsec:charac-advanced-attacks}
MTD techniques have been developed in order to deal with increasingly more sophisticated, intelligent, and persistent attacks equipped with more advanced tools. We discuss the key characteristics of these advanced attackers as follows:
\begin{itemize}
\item {\bf Persistent attackers}: In the existing MTD works, we found that attackers are {\em persistent}, not performing an one-time attack (e.g., APT attacks~\cite{ben2016attacker, DeLoach2014, evans2011effectiveness, hu2015dynamic}).  This persistent attack behavior is well observed in multi-stage attacks that start from scanning attacks in the reconnaissance stage prior to attacks obtaining access into a system (i.e., outside attackers) and continue to the attack delivery and/or exploitation stage after they break into the system (i.e., inside attackers).
\item {\bf Adaptive attackers}: 
Attackers are adaptive to dynamically changing system conditions and external environmental conditions, as they take into consideration both physical and cyber accessibility. These attackers also have intelligence with regard to their resources, executing adaptive attacks~\cite{Jia:CatchMeDSN2014, wang2014moving, wright2016moving} that wisely manage their resource limits and at the same time opportunistically seek to compromise an entire system.
\item {\bf Stealthy attackers}: Attackers do not exhibit an identifiable attacking behavior all the time. They perform attacks in a highly stealthy manner~\cite{Al-Shaer13, Jafarian:OFRHM2012}, even showing well-behaved features of good citizens. However, at a time 
when an attack is calculated to inflict serious harm or damage to the system, they exhibit attack behaviors. But they stay stealthy until the time comes. 
\item {\bf Incentive-driven attackers}: MTD is developed to deal with smart attackers. In particular, attackers are intelligent enough to execute attacks efficiently, such that the attack has a minimal cost but maximal outcome. Therefore, we can consider the attacker to be a rational actor that is sensitive to incentives, such as attack success with minimum cost~\cite{feng2017signaling}. 
\end{itemize}
As a typical type of sophisticated attacks, the {\em Advanced Persistent Threat} (APT) has been commonly considered as an attack type techniques need to countermeasure~\cite{ben2016attacker, DeLoach2014, evans2011effectiveness, hu2015dynamic, prakash2015empirical, zhu2013game}. More or less, APT attackers show the above features in performing their attacks and are often described under the scenario of the Cyber Kill Chains~\cite{hutchins2011intelligence}.  In the next section, we discuss an overview of the cyber kill chain and the behavior of an APT attack at each stage of the chain so as to understand how MTD techniques needs to consider APT attacks.

\subsection{Cyber Kill Chain} \label{subsec:ckc}
A new class of adversaries with sufficient resources and high intelligence has emerged, which is the so-called `Advanced Persistent Threat (APT),' as mentioned in Section~\ref{subsec:charac-advanced-attacks}. APT is capable of multi-year intrusion campaigns to obtain highly sensitive information, such as corporate proprietary information or national security secrets~\cite{hutchins2011intelligence}.  In order to defend against APTs, an intelligence-based defense model is critical to defenders for mitigating risk and/or vulnerability of attacks to a system. \citet{hutchins2011intelligence} developed the intelligence-based defense model, called the {\em Cyber Kill Chain} (CKC), based on multiple phases of a cyber attack. The CKC model provides: (1) the description of intrusion phases; (2) mapping the indicators of adversary kill chain to the courses of defense actions; (3) the pattern identification associating individual intrusions with broader campaigns; and (4) the understanding of iterative intelligence gathering. 

The CKC model consists of the following phases~\cite{hutchins2011intelligence}:
\begin{enumerate}
\item {\em Reconnaissance}: This phase involves researching, identifying, or selecting targets, often by crawling Internet websites to obtain target information, such as email addresses, social relationships, or information regarding particular technologies the target system uses.
\item {\em Weaponization}: This phase uses a remote access trojan to be coupled with an exploit into a deliverable payload (e.g., Adobe Portable Document Format or Microsoft Office documents) using an automated tool, called a {\em weaponizer}.
\item {\em Delivery}: In this phase, the weapon is transmitted to a targeted system using delivery vectors for weaponized payload by APT actors (e.g., email attachments, website, and/or USB removable media).
\item {\em Exploitation}: After the delivery of the weapon to a victim host, the intruders' code is triggered by exploitation, commonly targeting for an application or OS vulnerabilities.
\item {\em Installation}: The adversary can stay inside the target environment by installing a remote access trojan or backdoor on the target (or victim) system.
\item {\em Command \& Control} (C2): Compromised hosts try to establish a C2 channel by beaconing outbound to an Internet controller server. Upon the establishment of the C2 channel, the intruders can take control of the target system (e.g., `hands on the keyboard').
\item {\em Actions on Objectives}: After the previous six phases are successfully performed, adversaries launch attacks to breach security goals (e.g., data integrity, availability, confidentiality).
\end{enumerate}
MIT Lincoln lab used a shorter version of the CKC model with the phases of reconnaissance, access, exploit development, attack launch, and persistence~\cite{okhravi2013survey, Ward18-survey}. 
In this MIT Lincoln Lab version, the `access' phase incorporates the `weaponization' and 'delivery' phases from \cite{hutchins2011intelligence} and the `attack launch' phase incorporates `installation' and `C2' \cite{hutchins2011intelligence} as well.

\vspace{1mm}
\noindent {\bf Limitations of Considering APT attacks:} Existing MTD approaches heavily consider attackers in the stage of reconnaissance when the attackers are outside attackers. However, in many cases, stealthy, undetected inside attackers are more serious threats to the system when they are not properly handled, such as detecting them by an IDS. But considering a high volume of false detection, MTD can significantly help intrusion detection by adding another layer of defense against the inside attackers.

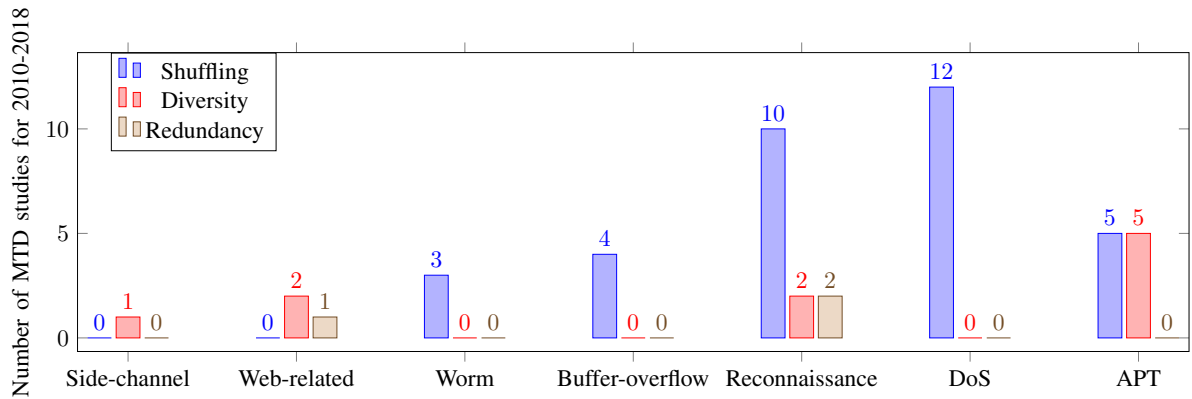
\begin{figure*}
    \centering
\begin{tikzpicture}[scale=0.9]
\begin{axis}[
    ybar,
    ymax=13,
    enlargelimits=0.05,
    legend style={at={(0.03, 1)},
      anchor=north west,legend columns=1},
    ylabel={Number of MTD studies for 2010-2018},
    symbolic x coords={Side-channel, Web-related, Worm, Buffer-overflow, Reconnaissance, DoS, APT},
    xtick=data,
    nodes near coords,
    nodes near coords align={vertical},
    ]
\addplot coordinates {(Side-channel, 0) (Web-related, 0) (Worm, 3) (Buffer-overflow, 4) (Reconnaissance, 10) (DoS, 12) (APT, 5)};
\addplot coordinates {(Side-channel, 1) (Web-related, 2) (Worm, 0) (Buffer-overflow, 0) (Reconnaissance, 2) (DoS, 0) (APT, 5)};
\addplot coordinates {(Side-channel, 0) (Web-related, 1) (Worm, 0) (Buffer-overflow, 0) (Reconnaissance, 2) (DoS, 0) (APT, 0)};
\legend{Shuffling, Diversity, Redundancy}
\end{axis}
\end{tikzpicture}
    \caption{Different Attacks considered by shuffling, diversity, or redundancy-based MTD approaches.}
    \label{fig:attack_types}
\end{figure*}

\subsection{Attack Types} \label{subsec:attack-types}
In this section, we discuss more concrete examples of attack behaviors considered in the existing MTD approaches. 

\begin{itemize}
	\item {\bf Reconnaissance (or scanning) attacks}: Scanning attacks are used by attackers to gather information and intelligence about a target system before an actual attack is launched. The attackers usually use a customized set of software tools (e.g., {\em Nmap}~\cite{Lyon:Nmap2009}) to scan the target system (or network) to find information, such as OS types, IP addresses, port numbers, running services, protocols, network topology, and exploitable vulnerabilities. An attacker/scanner performs scanning by sending probe packets (e.g., ICMP echo request message or TCP SYN, TCP ACK, TCP Xmas, UDP port scanning, etc.) to the target host(s) in the networks. 
	To counter reconnaissance attacks, several network address-based approaches have been proposed~\cite{Achleitner:2017Deceiving, Al-Shaer13, Carroll14, Duan:RRM2013, Jafarian:OFRHM2012, Kewley:DyNAT2001, Luo:RPAH2015, Sharma18}.
	\item {\bf DoS (or DDoS) attacks}: A DDoS attack uses a large number of botnets (i.e., a set of bots, consisting of compromised machines, or zombie machines) to attack a targeted system by flooding traffic messages (e.g., UDP floods, ICMP floods, and/or SYN floods) from multiple sources to force it to shut down or to deny services to legitimate users. DDoS often involve time-based attacks that can cause targeted systems to miss critical deadlines for updates or to fail to complete critical tasks. Common MTD approaches to handle these attacks include the use of hidden proxies, IP/port shuffling and/or address mapping~\cite{Aydeger:2016Mitigating, clark2015game, Ge14, Groat12, Jia:CatchMeDSN2014, jia2013motag, Kewley:DyNAT2001, Lee:PortHopping2004, Shi:PortAddressHopp2007, Steinberger:2018DDoS, wright2016moving, meier2018nethide}. 
	\new{For example, \citet{meier2018nethide} proposed a novel network obfuscation approach and implemented a framework named `NetHide' which can successfully battle against the possible attacks such as Link-Flooding Attacks (LFAs) launched by even advanced attackers. The MTD strategy behind the NetHide is to change and modify path tracing probes in the data plan. They showed that NetHide can hold a trade-off between security (i.e., creating difficulties for attackers) and usability (e.g., high accuracy and low network functionality degradation). Their results showed that NetHide can mitigate the probability of attack success by 1\% while it provides 90\% and 72\% accuracy and utility, respectively.} Moreover, MOTAG~\cite{jia2013motag} mitigates the flooding DDoS attacks using hidden proxies as moving targets to secure service access for authenticated clients. \item {\bf Buffer overflow attacks}: This kind of attack usually happens due to a lack of buffer boundary checking. Common memory protection techniques, such as Address Space-Layout Randomization (ASLR)~\cite{Shacham:ASR2004}, and Address Space Layout Permutation (ASLP)~\cite{Kil:ASLP2006}, are proposed to thwart the buffer overflow attack. These techniques randomize the memory position of data and program segment, library, and so forth. \citet{manadhata2013game} and \citet{Manadhata11-2} also considered a buffer overflow, leveraging the attack surface of a defense system and proposed a countermeasure against it by reducing or shifting the attack surface in terms of the system's methods, channels, and/or data items.
	\item {\bf Worm attacks}: A worm is a malicious software (malware) that replicates itself and employs a network to spread itself to other machines by relying on existing bugs or holes in the target system. The worms cause damages, such as consuming bandwidth or turning a worm-infected computer into a botnet that can be used by 
	the worm's programmer to profit by sending spam or conducting DDoS attacks.  A countermeasure against worm attacks is Network Address Space Randomization (NASR)~\cite{Al-Shaer13, antonatos2007defending,  Jafarian:SpatioMutation2014} that randomizes the host IP address (i.e., IP shuffling).
	\item {\bf APT attacks}: APT attacks are well known as one of the sophisticated attack types aiming to access system resources, control them, and perform exfiltration attacks (e.g., leaking sensitive or confidential information out to unauthorized parties) by performing stealthy, persistent, and adaptive attacks~\cite{Achleitner:2017Deceiving, Anderson16, carter2014quantitative, carter2014game, Cho18, feng2017signaling}. The behaviors of APT attackers can be observed in the cyber kill chain, which is described in Section~\ref{subsec:ckc}. APT attacks are also called `multi-stage attacks' that refer to attackers performing multiple attacks across multiple stages. For example, the attacks range from network scanning and packet sniffing to illegitimate authentication and service interruption (e.g., Stuxnet virus)~\cite{ben2016attacker, DeLoach2014, evans2011effectiveness, prakash2015empirical, zhu2013game}.	
	\item {\bf Side channel attacks on VMs}: Attackers perform attacks over side channels based on the shared CPU cache of a VM in order to obtain sensitive information~\cite{zhang2012incentive}.
	\item {\bf Attacks on Web applications}: Diverse types of attacks can be performed on web applications, such as SQL injection, directory traversal, and cross site scripting which breach the key security goals that are confidentiality, integrity, and/or availability. These types of attacks can be countermeasured by software stack diversity and/or redundancy~\cite{neti2012software, sengupta2017game, yuan2013architecture}.	
\end{itemize}
We summarized the main attack types considered in existing MTD approaches with 36 papers published for 2010-2018 in Fig.~\ref{fig:attack_types}.  Note that more than one technique or attack can be considered in each work. The three major attacks the existing MTD techniques have mainly addressed are reconnaissance attacks, DoS attacks, and APT attacks. In addition, most proposed approaches are shuffling-based MTD while diversity-based MTD techniques are often combined with the shuffling-based techniques. But redundancy-based techniques are rarely discussed to deal with the attack behaviors addressed above because the redundancy-based approaches have been studied in the dependability domain (e.g., reliability, availability) rather than for the purpose of MTD and often used in combination with either shuffling-based or diversity-based MTD techniques. 

\subsection{Limitations of the Current Attack Models}
From the survey of attack types considered by existing MTD approaches, we can identify the following limitations in the current attack models: 
\begin{itemize}
\item {\bf Smart, intelligent attackers are significantly less considered.} Most MTD techniques are shuffling-based to deal with attacks. If attackers are intelligent such that they are able to detect defense patterns by using learning mechanisms (e.g., machine learning or cognitive learning), then they can easily capture what types of MTD techniques are used and what the patterns are to trigger an MTD operation. However, highly intelligent, learning attackers are not really 
considered. Generally, it is assumed that attackers have certain attack patterns, rather than that they learn and can launch adaptive attacks. Whereas the contrary is true for defenders; they are often permitted to take proactive, adaptive defense actions based on an assumed highly intelligent learning capability. This assumption may not be true; even the attackers are often smarter than the defenders in practice~\cite{ISTR}. 
\item {\bf Few scenarios are considered to deal with multiple strategies by attackers and defenders.} A specific attack type can be easily mitigated or prevented by a particular MTD technique. However, if an attacker identifies an exploit that is not covered by the deployed MTD technique, the attacker can even use the exploit and successfully launch its attack to penetrate into the system. Unfortunately, most MTD approaches focus on a single or a small set of attacks, and few scenarios have considered multiple strategies that can be taken by attackers or defenders. For example, the attacker may need to decide which system vulnerability it needs to target while the defender may want to determine which defense mechanism needs to be used, where both parties aim to choose a cost-effective decision under resource-constrained, time-sensitive settings.
\item {\bf An attacker has been less considered as a rational decision maker with learning ability}. The smart attackers can leverage how MTD works and what side-effect the MTD can introduce. In addition, they can capture the defense patterns for an adaptive MTD which aims to realize a cost-effective MTD. However, attackers with learning capability and their mental models have been rarely studied while high intelligence with the defender and corresponding intelligent MTD (e.g., machine learning-based MTD) is easily assumed and has been proposed~\cite{colbaugh12, Tozer15, vikram2013nomad, Zhu14-reinforcement}.
\end{itemize}

\begin{table*}[t]
\centering
\caption{Key Modeling and Solution Techniques of MTD and Their Pros and Cons.} \label{tab:solution-modeling}
\vspace{-3mm}
\begin{tabular}{|P{2cm}|P{2.7cm}|P{4cm}|P{4cm}|P{3cm}|}
\hline
\textbf{Approach}                                        & \textbf{Technique}        & \textbf{Pros}                             & \textbf{Cons}            & \textbf{Ref.}                    \\ \hline
\multirow{3}{*}{\begin{tabular}[c]{@{}l@{}}Game  Theoretic \\ (GT) MTD \end{tabular}}            & General Game Theoretic Approach    & \multirow{3}{*}{\begin{tabular}[c]{@{}l@{}} Capability to effectively model \\ interactions between players; \\ identifying optimal solution under \\ complex interactions between \\ multiple parties \end{tabular}}                      & \multirow{3}{*}{\begin{tabular}[c]{@{}l@{}} Issues associated with players' \\ bounded rationality (or irrationality) \\ and/or misperception; high solution \\ complexity;\end{tabular}}                                & \cite{ carter2014quantitative, carter2014game, colbaugh12, Ge14,  jia2013motag, neti2012software, wright2016moving, zhu2013game} \\ \cline{2-2} \cline{5-5}                & Bayesian Stackelberg Game &                     &    & \cite{clark2015game, feng2017signaling, Paruchuri07,  Paruchuri08, sengupta2017game, zhang2012incentive, zhu2012deceptive}       \\ \cline{2-2} \cline{5-5} 
                       & Stochastic Game           &        &    & \cite{manadhata2013game, Shapley53}         \\ \hline
Genetic  Algorithm (GA)-based  MTD                       & General GA                & \begin{tabular}[c]{@{}l@{}}Fast method to find good solutions; \\ providing good  understanding of \\ how good a given solution is \\based on a fitness function;\end{tabular} & \begin{tabular}[c]{@{}l@{}} Scalability problem in a large \\ number of  generators; it may not \\ find the optimal solution; \\ uncertainty to find fitness\end{tabular} & \cite{crouse2011a,Crouse12ImprovingTD,John14,Zhuang:TheoryMTD2014} \\ \hline
\multirow{3}{*}{\begin{tabular}[c]{@{}l@{}}Machine  Learning\\ (ML)-based MTD\end{tabular}} & General ML & \multirow{3}{*}{\begin{tabular}[c]{@{}l@{}} Capturing an evolving attacker; \\ high scalability and applicability\end{tabular}}                                 & \multirow{3}{*}{\begin{tabular}[c]{@{}l@{}} Needing a large amount of data for \\ training and modeling\end{tabular}}                   &  
\cite{vikram2013nomad} \\ \cline{2-2} \cline{5-5} 
                    & Classification-based      &                     &                      & \cite{Colbaugh13, colbaugh12}                    \\ \cline{2-2} \cline{5-5} 
                    & Reinforcement learning    &    & &                 \cite{Tozer15, Zhu14-reinforcement}             \\ \hline
\end{tabular}
\end{table*}

\section{Modeling and Solution Techniques of MTD} \label{sec:modeling-solution-mtd}
MTD techniques have been proposed by using various types of modeling and solution techniques. In this work, we discuss key modeling and solution techniques of MTD based on the following theoretical approaches: (1) game theory; (2) genetic algorithms; and (3) machine learning. 

\subsection{Game Theory-based MTD} \label{subsec:game-modeling-solution}
The fundamental mechanism behind MTD techniques is to add another layer of defense by manipulating the attack surface that aims to increase the level of protection provided to a system while incurring additional cost (i.e., reconfiguration cost) and causing potential service unavailability to normal users. In this context, the clear concepts of {\em gain} and {\em loss} make the use of game theoretic approaches highly relevant to design and analyze cyberdefense as a game between an attacker and defender. Game theory can appropriately model decisions and actions by the defender and attacker, where they are assumed rational and aim to seek optimal strategies to maximize their utilities. Hence, in terms of the defender's perspective, the key goal of an MTD strategy is to identify a set of optimal system configuration policies in order to shift the attack surface, which can minimize risk and/or damage introduced by an action by the attacker~\cite{zhu2013game}. On the other hand, the attacker aims to successfully launch its attack with minimum effort / time and maximum effectiveness in achieving its objectives. Hence, a two-player game well models the competition scenario between the attacker and the MTD-based defender.

In this section, we discuss common game theoretic approaches used to develop MTD techniques, including a general game theoretic approach 
(i.e., not based on a specific game), a Bayesian Stackelberg Game, and a stochastic game. In addition, we discuss the cons and pros when 
game theory inspires the development of MTD techniques.

\subsubsection{General Game Theoretic Approaches}
Many MTD techniques have been developed using a general game framework where an attacker and defender are assumed to be rational and aim to maximize their respective utility. Although no specific game theory is mentioned, if authors say they take a game theoretic approach for their MTD by using the concept of payoff (or utility) based on estimated gain and loss in choosing the best strategy, in response to an action taken by its respective opponent, we categorized it as a game theoretic MTD approach.

\citet{zhu2013game} investigated a tradeoff between security and usability based on the security enhanced by the MTD and the performance degraded by the MTD, such as service unavailability or system reconfiguration cost. The authors modeled a game between an attacker and defender where the defender aims to minimize risk and maintain service availability while the attacker dynamically exploits vulnerabilities of system components with the goal of introducing a maximum damage on the system. The defender performs MTD by continuously changing their defensive strategies based on dynamically learned information in a highly uncertain environment.  \citet{Ge14} provided an incentive-compatible MTD technique based on server location migrations and a user-server mapping mechanism to enhance resilience and agility of a system along with high network timeliness and throughput.  \citet{carter2014quantitative, carter2014game} used a game theory to derive an optimal migration strategy. Those works analyzed temporal platform migration patterns and identified an optimal strategy for selecting the next platform. This demonstrates the following: (1) although platform diversity is effective against some persistent attacks, it may have a negative impact on the ability to defend against fast and local attacks~\cite{carter2014quantitative}; (2) increasing diversity in platform selection is more effective than randomization; and (3) ensuring high diversity between available platforms has a stronger impact on security than having a large number of available platforms. 

In contrast, \citet{colbaugh12} found that uniform randomization is an optimal strategy for diversity-based MTD. The differences between optimal strategies can be explained by the attack model each study used and more specifically whether the attacker's goal required a persistent foothold in the system. \citet{neti2012software} used an anti-coordination game to capture the interplay of choice, diversity, and scalability of risk in software-defined networking (SDN)-based MTD. This study evaluated a scenario where one node in a network is compromised while the rest of the nodes use a game theoretic approach to decide whether to switch or not per distinct platform. \citet{jia2013motag} proposed an approach to counteract DDoS attacks where the approach uses a pool of redundant proxies between the clients and the web application. To evaluate this concept, \citet{wright2016moving} conducted a game theoretic analysis where two players try to influence the quality of service experienced by users of a web application, at a minimal cost. Their investigation demonstrated how the effectiveness of mitigating DDoS strategies changes depending on various conditions (e.g., migration cost, a number of attacker bots, and/or information provided by an insider).

\subsubsection{Bayesian Stackelberg Game}

{\em Bayesian Stackelberg games}~\cite{Paruchuri07, Paruchuri08} have been used in modeling an attack-defense game to solve various cybersecurity problems. In a Stackelberg game, there are two players, a leader and a follower where the leader takes an action first and then the follower takes its action by investigating the impact of the leader's action to its payoff. Hence, the follower aims to optimize the payoff of its action based on the leader's action. Hence, in this context, what action to take by the leader is critical to leading the follower's action. The Stackelberg game has been popularly used in modeling interactions between an attacker and defender in a system with MTD techniques as explained below.

\new{\citet{hasan2017protection} proposed a game theoretic model using Nash equilibrium named Co-resident Attacks Mitigation and Prevention (CAMP) to detect co-resident attack and mitigate the malicious VM co-location in a co-resident environment. Through simulation, they showed that their game model can provide optimal defensive strategies for the VM which can effectively fail the co-resident attack.} \citet{feng2017signaling} investigated how the strategies of information disclosure by defenders can improve the effectiveness of MTD techniques based on Bayesian Stackelberg game theory. They designed a signaling game based on the concept of the Bayesian Persuasion Model to consider how a defender signals an attacker and how the attacker responds to the signal in their decision making process. Based on the analysis of the optimal defense strategies in platform migration, this work shows that strategic information disclosure is a promising method to enhance defense effectiveness.  \citet{zhang2012incentive} proposed an incentive-compatible MTD to identify an optimal interval of VM migration in clouds. This work uses the Vickrey-Clarke-Groves (VCG) mechanism to realize {\em mechanism design} by considering the security benefit and the defense cost introduced by the VM migration. 

Some shuffling-based MTD techniques are used to implement defensive deception techniques. These types of techniques can be called `MTD using deception.' For example, \citet{clark2015game} used IP mutation to distract and/or mislead attackers to fail to identify real nodes. This work places a network of decoy nodes and identifies an optimal strategy to mutate IP addresses of real nodes based on a Stackelberg game.  \citet{zhu2012deceptive} proposed a deceptive flow-based defense mechanism in a multi-path routing network by generating fake packets aiming to lure attackers to expend their energy disrupting fake packets. This work modeled the interactions between an attacker and a defender based on a Stackelberg game and derived solution equilibria based on an iterative backward induction method. 

\citet{sengupta2017game} used a Bayesian Stackelberg game to identify an effective switching strategy for web applications to maximize security for a given set of system configurations while minimizing the defender's switching cost. 

\subsubsection{Stochastic Game}
{\em Stochastic game} is developed by \citet{Shapley53} and it can reflect dynamics between multiple players based on probabilistic transitions.  The game consists of multiple stages where each stage can start from some state. Each player chooses its action and receives its respective payoff based on the taken action and the current state. The process is repeated as players arrive new states based on the probabilistic transitions.  \citet{manadhata2013game} proposed a two-player stochastic game model to determine an optimal MTD strategy based on attack surface diversification. The authors explicitly modeled different attacker profiles (e.g., script kiddies, experienced hackers, organized criminals, and nation states) and used the subgame perfect equilibrium concept to determine the optimal defense strategy. 

\vspace{1mm}
\noindent {\bf Pros and Cons}: 
Since game theoretic approaches have been so commonly used in modeling the competitive interactions between an attacker and defender in designing techniques for MTD, we discuss several key advantages and disadvantages of game theoretic MTD.

{\bf The advantages of game theoretic approaches} are:
\begin{itemize}
\item {\bf Game theory offers an effective way to formulate interactions between an attacker and defender.} Game theory has been well explored to model the decision making process and how the best strategy is chosen by multiple players who participate in the same game. Hence, there are many existing, mature game theories to model different scenarios of a cybergame by an attacker and defender. In particular, game theoretic modeling techniques are highly effective to model actions by each party based on historical interactions between them. 
\item {\bf Game theory provides convenient tools to formulate the decision utility of an attacker and defender.} Most cybergame scenarios consider a player's multiple objectives. For example, the attacker wants to maximize their attack effectiveness (i.e., impact by the attack success) with the minimum attack cost (e.g., time / resource to perform an attack). In addition, the defender also chooses a defense strategy to maximize defense strength for enhanced system security while minimizing service interruptions to users and defense cost. Both parties basically want to take an action that provides the best outcome with minimum cost.  
\item {\bf Game theory can provide an optimal strategy based on learning.}  We can easily embed each player's learning toward an opponent's action into a game. In particular, in a cyberwar game with the key players being an attacker and defender, taking an adaptive strategy under a highly dynamic, hostile environment is critical to achieving their respective goal. Game theoretic approaches can provide an effective way to consider each party's learning and accordingly their adaptive behavior to maximize their utility under the dynamic, hostile settings. 
\end{itemize}

\noindent {\bf The disadvantages of game theoretic MTD} are:
\begin{itemize}
\item {\bf Attackers are not necessarily rational or intelligent.} Game theory assumes that all players are rational in order to maximize their own utilities. However, in the context of cybersecurity, attackers may not necessarily be rational or intelligent. Instead, the volume of attackers can be substantially large even if their intelligence is very low, aiming only to waste resources of a system (or network) and decrease the system resilience, leading to a sudden breakdown of the system without recoverability. In addition, if incentives based on rationality are not compatible (e.g., attackers are not stimulated by the incentives), the best strategies derived by the defender may not work for irrational attackers.
\item {\bf Solution space may be too large, resulting in prohibitively high solution complexity.} A game theoretic approach can provide a best strategy for a defender to mitigate the impact by an attack. However, generating an optimal strategy may not be light enough to be run particularly on resource-constrained platforms. Greedy, heuristic approaches usually sacrifice a certain level of optimality while providing lightweight solutions.
\item {\bf Players, an attacker and a defender, may have inherent misperception impacting their decision making process.} As the assumption of complete information available to players has been realized as unrealistic, game theoretic approaches considering incomplete information have been proposed~\cite{Harsanyi95} meaning that players do not have perfect knowledge in deriving their accurate utilities. Nevertheless, since incomplete information is applicable to all players in a game, the view toward the game itself is assumed to be the same, implying that each player play the same game based on available information even if the information itself would be incomplete. But in reality, each player may interpret the same game differently based on its own subjective perception. Hypergame theory~\cite{Gharesifard10} has been proposed to deal with this kind of misperception and/or uncertainty problem in game theory; but it has not been applied in modeling interactions between an attacker and defender in a system with MTD.
\end{itemize}
\vspace{-2mm}
\subsection{Genetic Algorithm-based MTD} \label{subsec:meta-heuristic}
Genetic algorithms (GAs) have been used to develop MTD mechanisms. \citet{crouse2011a} employed a GA to identify a secure computer configuration (e.g., OSes or applications) with high diversity in time and/or space. They modeled a computer configuration as a chromosome in which an individual configuration is treated as a trait or allele. Based on the attack resilience, a top ranked computer configuration is selected. The authors extended their work in \cite{Crouse12ImprovingTD} to investigate how `chromosome pool management' can enhance the diversity of computer configurations based on a GA-based approach. In particular, they considered the aging aspect of configurations to reflect vulnerability that can be introduced by an aged configuration. In addition, the authors enhanced their GA-based MTD to change computer configurations by changing mutation as well as using the feedback about system security status~\cite{John14}. \citet{Zhuang:TheoryMTD2014} also used a GA to generate system configurations with high diversity for maximizing system security.

\new{\citet{Lucas:EvolMTDFramework2014} described a host-level  implementation of an evolutionary strategy for MTD that proactively discovers secured alternative configurations over time. This work is based on the evolutionary approach originally introduced in~\cite{crouse2011a}. The evolutionary based MTD models computers as chromosomes. The computer configurations are generated using an evolutionary algorithm (EA) with a series of operations such as reproduction, recombination, and mutation. This implementation prototype consists of mainly the following three components: (i) EA that discovers the configurations; (ii) virtual machine (VM) that implements the algorithm; and (iii) assessment for scoring the chromosomes.}

\new{Recently,~\citet{Ge:2019modeling} discussed a GA-based approach for optimizing the network shuffling in an SDN-based IoT network. In this work, the authors considered two types of the IoT nodes (e.g., decoy and real nodes), and designed three metrics, including the number of attack paths towards the decoy targets, mean time to security failure, and defense cost. These metrics are optimized for the network shuffling technique.
}

\vspace{1mm}
\noindent {\bf Pros and Cons}: GAs can be a useful approach as they attempt to find the best or near optimal solution(s). However, developing a fitness function that can provide a diverse solution space for design decisions in MTD techniques is not a trivial task as the design of an MTD technique 
needs to accommodate multiple conflicting system goals.  In addition, for a large solution space, efficiency becomes an issue, limiting scalability. Further, for resource-constrained environments, a GA-based solution is not attractive due to high complexity. 
\vspace{-2mm}
\subsection{Machine Learning-based MTD}
\citet{vikram2013nomad} proposed an MTD technique by randomizing HTML elements for web services to deal with the web bot attacks.  The authors used ML to enhance the effectiveness of the proposed strategy, evaluating their proposed technique by measuring page loading time overhead. This ML-based MTD technique incurs low overhead while effectively thwarting attacks.

\citet{colbaugh12} proposed a predictive MTD technique using ML aiming to mitigate the adversary's ability to learn about the defensive mechanism. In this work, the authors supposed that attackers can learn and may leverage a reverse-engineering method to anticipate the defensive strategies. They evaluated their algorithm using cybersecurity datasets to show the effectiveness and robustness of their approach. The authors used another approach to deal with the same problem in \cite{Colbaugh13}. They leveraged a defensive MTD and an ML-based method using the co-evolutionary relationship between both an attacker and  defender to derive an optimal defensive strategy that is hard to reverse-engineer.

\citet{Zhu14-reinforcement} proposed two iterative reinforcement learning algorithms to identify an ideal defense strategy against cyberattacks especially when the information about the attackers is unknown or limited. They used Markov chains and stochastic stability in the algorithms by introducing the adaptive, robust reinforcement learning capability. They showed that their method can provide the nearly optimal defensive strategy. \citet{Tozer15} proposed a multi-objective reinforcement learning algorithm to minimize the attack surface of a system. They designed a system to generate a multi-objective Markov Decision Process using the system's components and behaviors to identify optimal policies.

\new{\citet{Sengupta:MTDeep2017} proposed an MTD framework (MTDeep) for Deep Neural Networks (DNNs) that increases security and robustness of the DNNs against the adversarial attacks. In MTDeep, an input image is classified randomly selecting a network from an ensemble of the networks which is based on a strategy generated via game-theoretic reasoning. An interaction between the image classification
system is modeled with MTDeep (i.e.,  an ensemble of DNNs) and its users (i.e., adversarial and legitimate) into a Repeated Bayesian Game. The defender's configurations space are the ensemble of DNNs which are trained for the same task, but they are not affected by the same attack. Stackelberg equilibrium of the game provides the optimal switching strategy for MTDeep reducing the misclassification on adversarially modified image with high classification accuracy for the legitimate users of the system. 
Recently, \citet{Song:DeepVisual} designed an MTD-based approach for embedded deep visual sensing systems against adversarial examples for generating multiple new deep models (e.g., multilayer
perceptrons and convolutional neural networks) which can be used to collaboratively detecting and thwarting the adversarial examples.  The adversarial examples are inputs to a neural network (e.g., deep model) that makes wrong classification results.  The deep models are generated dynamically using the concept of MTD after the system deployment. The post-deployment of the models are different across the systems. This approach invalidates an essential basis to the attackers and disrupts them to build effective adversarial examples. Similarly, \citet{Farchi:MLSelection} proposed a strategic ML-selection approach defending against the adversarial machine learning. In this work, the authors suggested that the attacks against the learning can be reduced with careful design of strategic selection of learning methods and attributes. The defender implements multiple learners with their strategic activation computation using the game-theoretic approach. 
}
 
\vspace{1mm}
\noindent {\bf Pros and Cons}: 
ML-based MTD allows a system to capture evolving attack patterns with high scalability and applicability. However, as the performance of ML often requires a large amount of data for training in order to guarantee a certain level of prediction accuracy, then when there is a lack of data, the performance is less than desired even with high overhead and complexity.  Further, we need to ensure a sufficient level of computational power available in a given environment where MTD is deployed as some resource-constrained environments cannot afford ML-based MTD.

Some other MTD techniques have used other approaches, which do not use any of the above methodologies (i.e., game theory, genetic algorithms, or machine learning). Those are already discussed in Section~\ref{sec:mtd_techniques} based on the three categories of operation-based MTD (i.e., shuffling, diversity, and redundancy).

\subsection{Discussions on the Existing MTD Modeling \& Solution Techniques}

We observed the following trends in key modeling and solution techniques used to develop MTD in the literature:
\begin{itemize}
    \item {\bf A large volume of game theoretic MTD approaches}: As discussed earlier, due to the substantial advantages of game theoretic approaches in terms of the flexibility of problem formulation that can reflect diverse scenarios in most domains, game theoretic approaches are dominantly employed in developing MTD. However, again we should keep the caveats in mind in terms of the rationality of players, solution complexity, and misperception or uncertainty in decision making, as discussed in Section~\ref{subsec:game-modeling-solution}. In addition, we found that several game theories are dominantly used such as a Bayesian Stackelberg game or a stochastic game. To better deal with uncertainty or misperception in the decision making, hypergame theory~\cite{Gharesifard10} can be applied to consider more realistic scenarios which should reflect uncertainty and bounded rationality.  
    \item {\bf GA-based MTD for identifying the optimal deployment of system configurations}: When a problem size becomes large, evolutionary algorithms, such as genetic algorithms (GAs), are often used to maximize solution optimality. But as discussed earlier, the downside of GA-based MTD is a lack of scalability with high complexity for large-scale or resource-constrained environments. In addition, ensuring a centralized entity to make MTD decisions based on GAs may not be guaranteed for some fully distributed environments.
    \item {\bf ML-based MTD to hinder an attacker's learning or select a best defense strategy}: The high performance of ML in learning attack patterns or identifying optimal solutions is attractive when a large volume of data is available. However, under a small set of data and/or highly dynamic datasets, ML-based solutions can be somewhat expensive without introducing a significant benefit of using it. Further, we need to ensure a sufficient level of computational power available by a trusted entity to manage MTD operations. 
\end{itemize}
Table~\ref{tab:solution-modeling} summarizes the key modeling and solution techniques of MTD and their pros and cons discussed above.

\section{Metrics for MTD} \label{sec:metrics}
The underlying idea of MTD techniques has been explored; accordingly many MTD techniques have been developed. However, no standard metrics have been proposed to measure their effectiveness and/or efficiency. In this section, we discuss what metrics have been used to assess the effectiveness and efficiency of the existing MTD techniques. Along with the discussions of the overall trends observed from this survey, we address some limitations of the metrics used in the state-of-the-art MTD approaches.

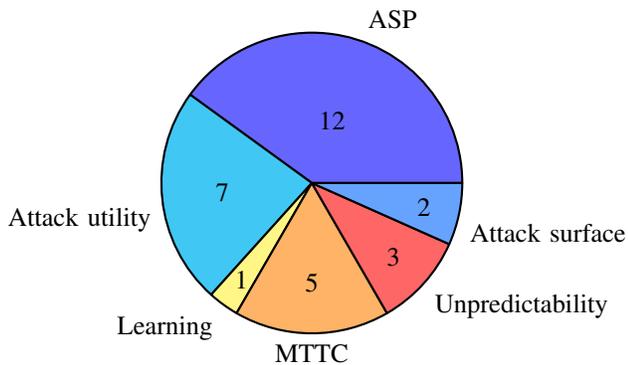
\begin{figure}
\centering
\begin{tikzpicture}[scale=1]
\pie[sum=auto , after  number=, radius =2]
{12/ASP, 
7/Attack utility,
1/Learning,
5/MTTC,
3/Unpredictability,
2/Attack surface
}
\end{tikzpicture}
\caption{Metrics measuring MTD effectiveness by an attacker's perspective.}
\label{fig:metrics-attack-effectivebess}
\end{figure}

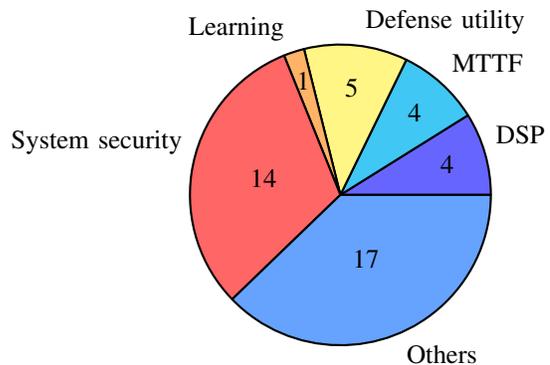
\begin{figure}
\centering
\begin{tikzpicture}[scale=1]
\pie[sum=auto , after  number=, radius =2]
{4/DSP,
4/MTTF,
5/Defense utility,
1/Learning,
14/System security,
17/Others},
\end{tikzpicture}
    \caption{Metrics measuring MTD effectiveness by a defender's perspective.}
    \label{fig:metrics-defense-effectivebess}
\end{figure}
\vspace{-3mm}
\subsection{Metrics for Measuring MTD Effectiveness} 
We discuss the metrics to measure the effectiveness of MTD techniques in terms of the perspectives of an attacker and a defender. The attacker's metric estimates its attack performance, indicating that the attacker's high performance refers to the defender's low performance, and vice-versa. The defender's metric measures its performance in achieving security and/or defense goals of a given system.

The {\bf attacker's metrics} estimates the adverse impact of the proposed MTD techniques on the attacker's performance and are obtained by:
\begin{itemize} 
\item {\bf Attack success probability (ASP)}~\cite{Al-Shaer13, Anderson16, ben2016attacker, Carroll14, carter2014quantitative, Casola2013IRI, Cho18, DeLoach2014, evans2011effectiveness, Rahman14, Sharma18, Zaffarano:Quantative2015}: This metric refers to the probability that attacks are successfully performed. For example, it refers to the probability that a system component (or defender) is compromised or a target is successfully discovered and/or accessed by an attacker. \citet{Rahman14} also used a metric called {\em attackability}, which refers to the probability that an attacker can access system states (or components) to the attack. This metric combines the degree of system vulnerability plus the feasibility an attacker accesses and performs an attack based on its own resource level. But in the sense that it measures the degree for the attacker to successfully access a system component, this metric is aligned with ASP.
\item {\bf Attack utility}~\cite{Alavizadeh18, alavizadeh2018evaluation, ben2016attacker, neti2012software, prakash2015empirical, zhu2013game, zhu2012deceptive}: When the interactions between attackers and defenders and their best strategies are considered based on game theoretic approaches, the payoff (or utility) of an attacker is used to measure the gain or loss by deploying a proposed MTD.  
\item {\bf Learning by attackers}~\cite{zhu2013game}: This measures the degree of an attacker's learning toward the payoff obtained by a defender upon the performed attack.
\item {\bf Mean time to compromise a system  (MTTC)}~\cite{alavizadeh2017effective, carter2014quantitative, carter2014game, Cho18, zhuang2014model}: This indicates how long an attacker takes to compromise an entire system. In terms of a defender's perspective, this metric is similar to {\em mean time to failure} (MTTF), as described under a defender's metrics below.
\item {\bf Unpredictability}~\cite{Green15, MacFarland15, shan2015proactive}: This indicates how much confusion and/or uncertainty a given MTD has introduced to attackers.
\item {\bf Attack surface}~\cite{Manadhata11-2, Manadhata11}: This metric is defined as the amount of system resources that can be used by attackers to attack the system, such as channels, data items, and/or methods. Hence, a larger attack surface exposes more vulnerabilities.
\end{itemize}

In the literature, various types of the {\bf defender's metrics} are used to measure MTD effectiveness in terms of the following metrics:
\begin{itemize}
\item {\bf Defense success probability (DSP)}~\cite{Al-Shaer13, Clark13, colbaugh12, Zaffarano:Quantative2015}: We call metrics measuring the success of an MTD technique DSP in this work. \citet{Zaffarano:Quantative2015} estimated the rate of executing successful defenses (e.g., for a defender, the rate at which tasks are executed and completed) or attacks (e.g., for an attacker, the rate at which attacks are performed and successfully completed). \citet{colbaugh12} used a detection accuracy of anomaly behaviors, such as attack behaviors or spams in order to determine whether to trigger an MTD operation. \citet{Clark13} measured the portion of decoy nodes detected by attackers when IP randomization techniques are used in order to measure the success of the IP-shuffling MTD strategy. \citet{Al-Shaer13} measured an IP-mutation success probability to measure the effectiveness of the MTD. This metric measures the probability that a mutated IP is not hit by scanning attacks.
\item {\bf Mean time to failure (MTTF)}~\cite{carter2014quantitative, carter2014game, Cho18, zhuang2014model}: This refers to a system reliability metric capturing the system's up-time in the presence of attacks when failures can happen due to either defects or security threats. This metric is the same as MTTC under the attacker's metrics.
\item {\bf Defense utility}~\cite{ben2016attacker, neti2012software, prakash2015empirical, zhu2013game, zhu2012deceptive}: Game theoretic approaches for the optimal deployment of MTD techniques have taken to identify the best defense strategy by a defender. The payoff (or utility) of the defender measures the effectiveness of an MTD technique.  
\item {\bf Learning by defenders}~\cite{zhu2013game}: This metric measures the degree of a defender's learning toward the payoff an attacker has obtained upon a defense action taken by the defender.
\item {\bf System security}: Various kinds of metrics measure the system security properties enhanced by proposed MTD techniques:
\begin{itemize}
\item {\bf Confidentiality}~\cite{prakash2015empirical, zhang2012incentive, Zaffarano:Quantative2015}: This measures how many system components are compromised~\cite{prakash2015empirical}. In some context, some information should be kept confidential, such as private information. The degree of preserving confidential or private information is another metric to indicate the degree of security. In~\cite{Zaffarano:Quantative2015}, mission confidentiality refers to the degree of exposing confidential information to unauthorized parties while attack confidentiality means the degree of attack behaviors detected by a defender.
\item {\bf Integrity}~\cite{Zaffarano:Quantative2015}: Integrity metric is discussed in terms of mission integrity and attack integrity. Mission integrity refers to how much information related to executing a given mission is communicated without being modified and/or forged, while attack integrity indicates how much accurate information the attackers view.
\item {\bf Availability}~\cite{Green15, prakash2015empirical}: This indicates the portion of system assets that are not compromised to provide a normal service.
\item {\bf Degree of vulnerability}~\cite{alavizadeh2017effective, Alavizadeh18, alavizadeh2018evaluation, ben2016attacker, Carroll14, carter2014quantitative, carter2014game, zhu2013game}: This measures the probability that a given platform to be selected is vulnerable during a particular time period or a given system component is vulnerable because it is controlled by an attacker.
\end{itemize}
\item {\bf Other metrics}: Based on the unique features of each of the existing MTD approaches, various other types of metrics have been adopted to measure the effectiveness of MTD as follows:
\begin{itemize}
\item {\bf Controllability}~\cite{prakash2015empirical}: This refers to the portion of critical system assets which expose a high vulnerability to an attacker if compromised.
\item {\bf Worm propagation speed}~\cite{Al-Shaer13, Jafarian:AddMutation2015, Jafarian:OFRHM2012}: This measures how much a deployed MTD can slow down actions by an attacker. This also indirectly increases the detection of attackers by earning more time to monitor the attacker.
\item {\bf Vastness}~\cite{Green15, MacFarland15, Manadhata11, zhang2012incentive}: This measures the size of spaces that a given defense mechanism can cover, such as IP spaces an attacker needs to scan through. In addition, the number of target hosts set by a defender can consume an attacker's resource because it determines all the possibilities the attacker needs to scan through. In~\cite{Manadhata11}, this is considered based on the metric called `attack surface measurements.'
\item {\bf Periodicity}~\cite{Green15, MacFarland15}: This estimates how often system configurations change in order to provide a sufficient level of confusion to attackers.
\item {\bf Uniqueness}~\cite{Green15, MacFarland15}: This measures how uniquely an individual entity (e.g., a host) is authorized to a system without being accessed by other entities.
\item {\bf Revocability}~\cite{Green15, MacFarland15}: This measures the degree of frequency to terminate or expire a prior system configuration (e.g., access control or a given IP configuration).
\item {\bf Distinguishability}~\cite{Green15, MacFarland15}: This measures how well a given defense distinguishes trustworthy entities from non-trustworthy entities.  
\item {\bf Loss in rewards between an optimal deployment and an executed deployment}~\cite{sengupta2017game}: This metric captures how much loss occurred for the actual execution of an MTD operation over the optimal deployment.
\end{itemize}
\end{itemize}
The metrics to measure the effectiveness of existing MTD techniques are summarized in Figs.~\ref{fig:metrics-attack-effectivebess} and \ref{fig:metrics-defense-effectivebess} based on 27 papers published during 2011-2018. Note that more than one metric can appear in a single paper. The general trends observed from the survey are: (1) the attack success probability (e.g., whether an attacker achieved its goal of a launched attack such as finding a vulnerable target) is a dominant metric used for the effectiveness of MTD aiming to minimize this metric; (2) due to a large volume of game theoretic approaches in the state-of-the-art MTD techniques, the payoff or utility of attackers or defenders is also one of dominant metrics used in the existing MTD techniques; and (3) some system-level metrics measuring system vulnerability or reliability (e.g., degree of vulnerability, MTTF or mean time to compromise an entire system) in the presence of attacks are observed as major metrics used to measure MTD effectiveness.  

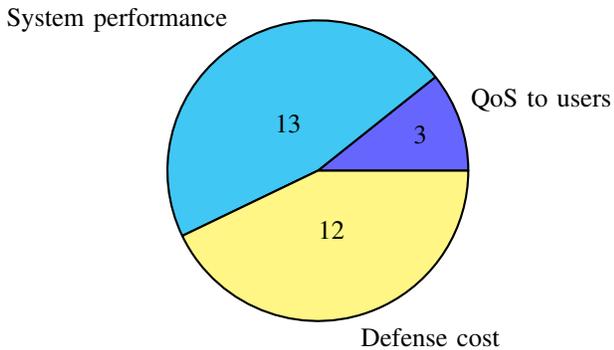
\begin{figure}[ht]
	\centering
\begin{tikzpicture}[scale=1]
\pie[sum=auto , after  number=, radius =2]
{3/QoS to users,
13/System performance,
12/Defense cost
}
\end{tikzpicture}
\caption{Metrics measuring MTD efficiency by a defender's perspective.}
	\label{fig:metrics_efficiency}
\end{figure}

\subsection{Metrics for Measuring MTD Efficiency} 

The {\bf attacker's metrics} are used to capture how much cost (or penalty) is introduced for an attacker to achieve attack success when a proposed MTD is deployed, as follows: 
\begin{itemize}
\item {\bf Penalty in attack payoff}~\cite{feng2017signaling, jia2013motag, wright2016moving, zhang2012incentive}: Many game theoretic MTD approaches estimate attack cost at an abstract level (e.g., cost is 1 for attacking; 0 otherwise).
\item {\bf Attack cost}~\cite{alavizadeh2018evaluation, Alavizadeh18, kampanakis2014sdn, manadhata2013game, Sharma18}: This measures how much overhead and/or impact is introduced to attackers to perform their attacks. To be specific, an attacker's scanning tool, such as $Nmap$, is used to capture the scanning overhead~\cite{kampanakis2014sdn}.
\end{itemize}
The MTD efficiency by the attacker's perspective is mainly measured by two types of metrics as above although both metrics are concerned about resources the attacker needs to invest to launch a planned attack. Due to the similar nature of both metrics and their small volume, we omit the figure.

The {\bf defender's metrics to measure MTD efficiency} that are commonly observed in the literature include:
\begin{itemize}
\item {\bf Quality-of-Service (QoS) to users}~\cite{Clark13, Han14, wright2016moving}: This metric captures the degree of service quality provided to users while implementing a given MTD technique as triggering an MTD (e.g., platform migration) often hinders service availability to normal users~\cite{Han14, wright2016moving}. In addition, upon deploying IP mutation techniques, the number of connections interrupted~\cite{Clark13} is used to measure QoS provided to users. 
\item {\bf System performance}~\cite{Albanese13, Chowdhary:2016SDN, clark2015game, Dunlop11, Ge14, Han14, Jackson13, Li14, Morrell14, Portokalidis:GlobalISR2011, vikram2013nomad, wang2014moving, Zhang17}: This metric measures how much overhead is introduced to deploy a given MTD, such as message overhead (e.g., delay, packet loss, or control packet overhead)~\cite{Dunlop11, Zhang17}, operational delay~\cite{Morrell14, Zhang17} / cost~\cite{Ge14, Han14, vikram2013nomad, wang2014moving, Zhang17} to deploy an MTD, the number of dropped connections~\cite{clark2015game}, or performance overhead (e.g., file sizes or performance degradation to distribute software for diversity)~\cite{Jackson13}. System performance is also captured by system throughput measuring how much a given system (or network) can maintain its performance in terms of network throughput (e.g., how many messages are correctly delivered)~\cite{Albanese13, Zhang17} or server throughput (e.g., how many queries are properly provided)~\cite{Portokalidis:GlobalISR2011}.
\item {\bf Defense cost}: Various aspects of defense cost are measured to indicate the efficiency of MTD techniques. An abstract level of defense cost (i.e., migration cost or maintenance cost of VMs)~\cite{Cho18, feng2017signaling,  jia2013motag, wright2016moving, zhang2012incentive} is used to measure the cost of a deployed MTD mechanism, which is mostly used in game theoretic MTD. Some defense cost captures the level of infrastructure (e.g., a number of proxy or decoy nodes) required to ensure a required level of service availability~\cite{jia2013motag, wright2016moving}. Some other works also used specific metrics to capture defense cost as follows:
\begin{itemize}
\item {\bf Address space overhead}~\cite{Al-Shaer13}: In deploying Random IP mutation techniques, this refers to the required address space based on mutation speed (e.g., low frequency mutation, LFM, or high frequency mutation, HFM).
\item {\bf Flow table size}~\cite{Jafarian:AddMutation2015, Jafarian:OFRHM2012, Zhang17}: This measures the size of flow table in OpenFlow (OF) switches when OF-RHM (Random Host Mutation) is used in an SDN-based MTD.
\item {\bf Integrated performance cost}~\cite{clark2015game, Ge14}: This metric integrates both performance and security cost. \citet{Ge14} considered bandwidth cost and risk at servers upon being attacked to calculate the overall performance cost. \citet{clark2015game} defined the cost function in terms of the number of active sessions, caused by triggering MTD operations, and the fraction of decoy nodes scanned by attackers. In these works, the goal is to minimize the performance cost that reflects defense cost, security, and service availability to users.
\item {\bf Strategy switching cost}~\cite{sengupta2017game}: This cost measures the switching cost (e.g., migration cost). \citet{sengupta2017game}  estimated the switching cost in switching web-stack configurations as a cost metric.
\item {\bf Power consumption}~\cite{Zeitz18}: When MTD is deployed in resource-constrained environments such as wireless sensor networks or IoT environments, energy consumption is one of the key design considerations. How much benefit is introduced over the power consumption by deploying an MTD technique is a critical metric to measure the efficiency of the MTD.
\end{itemize}
\end{itemize} 

The metrics to measure the efficiency of MTD by a defender's perspective are summarized in Fig.~\ref{fig:metrics_efficiency} based on 25 papers published during 2011-2018. Since more than one metric can be used in a paper, the number of metrics countered is not the same as the total number of works examined in this survey. 

As demonstrated in Fig.~\ref{fig:metrics_efficiency}, most metrics measuring MTD efficiency belong to system performance or defense cost. However, the level of QoS provided to users are significantly less studied. This means MTD technology focuses more on enhancing system security and performance with minimum cost while service availability for users to provide seamless, continuous service provision has remained much less explored in designing MTD techniques. Since deploying and executing MTD mechanisms to a system introduces a critical tradeoff between security, defense cost, and service availability, an optimization problem with these dual conflicting goals should be investigated in-depth to meet multiple criteria from both system goals (i.e., multi-objective optimization problem)~\cite{Cho17-moo}.

\section{Evaluation Methods for MTD} \label{sec: evaluation-methods}
MTD techniques have been verified by using various types of evaluation techniques. In this section, we discuss how the performance of MTD techniques have been assessed based on the following evaluation methods: (1) analytical models; (2) simulation models; (3) emulation models; and (4) real testbeds. 

\subsection{Analytical Model-based MTD Evaluation}

\subsubsection{Probabilistic Model-based MTD Evaluation}
In probabilistic models, the behaviors of a system, an attacker, and a defender and the interactions between them are described based on probabilistic parameters. \citet{Okhravi14} constructed a probability model to measure the mean time to security failure (MTTSF) where the security failure is defined by the system state being compromised by an attacker where the system is defended by MTDs. \citet{zhuang2012simulation}  also modeled the relationship between the frequency of diversity-based MTD and ASP. \citet{Carroll14} provided probabilistic models to measure the effectiveness of an address shuffling-based MTD technique based on ASP with respect to the network size, the addresses space scanned, the degree of system vulnerability, and the frequency of shuffling operations. 

\citet{Crouse:Prob2015} developed probabilistic models to measure ASP when a set of reconnaissance defenses, including honeypots as a deception technique and network address shuffling as MTD, is deployed in a given system, under varying the network size, the size of honeypots deployment, and the number of vulnerable nodes. \citet{Cho18} used a probabilistic model by building a Stochastic Petri Nets (SPN) model to describe an integrated defense system consisting of MTD, deception, and an IDS and analyzed the performance of the integrated defense system compared to the system with various combinations of defense mechanisms (i.e., an IDS only or IDS plus either deception or MTD) in terms of ASP and MTTSF (i.e., system lifetime). \citet{Sharma18} used probabilistic models to measure the effectiveness of the proposed IP-multiplexing based network shuffling techniques in terms of ASP and defense cost.
\citet{luo2014effectiveness} used probabilistic models to verify the effectiveness of a port hopping-based MTD technique against reconnaissance attacks in terms of ASP.

\vspace{1mm}
\noindent {\bf Pros and Cons:} The evaluation of probabilistic models provides insights and lessons based on the observations of general system behaviors with minimum evaluation cost. However, due to the probabilistic parameterization of all the key system design features, a certain level of simplicity and abstraction is unavoidable and may not be able to capture some deviations and/or unexpected effect that can be introduced in real application scenarios.  
\begin{figure}[t]
	\subfigure[An example cloud model.]
	{
		\centering
		\includegraphics[width=\linewidth,height=5.5cm]{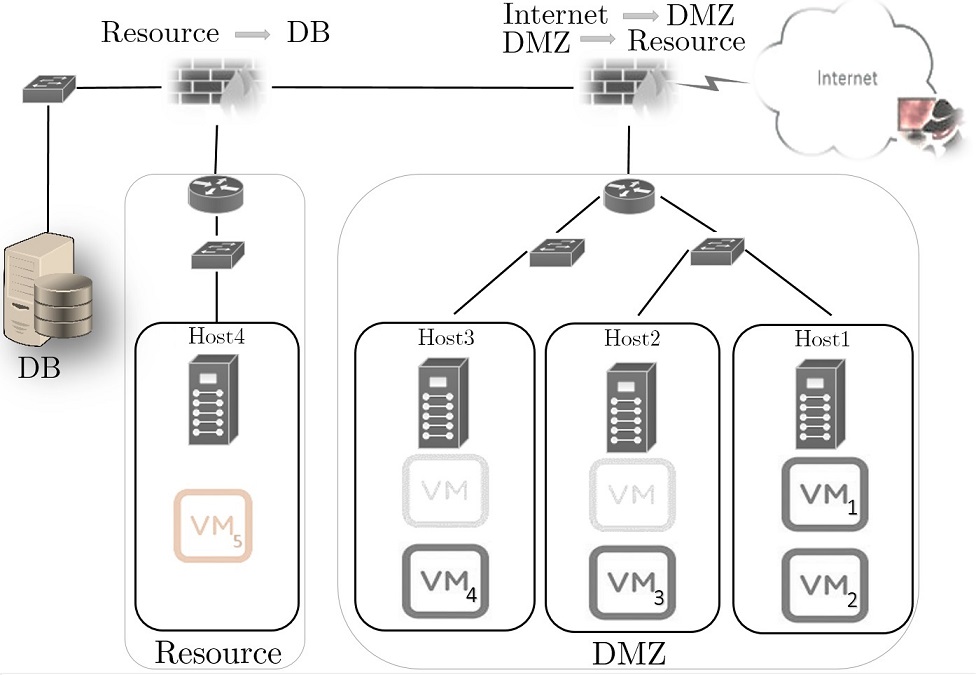} 
	}
	\subfigure[Two-layer HARM of the example cloud in (a).]
	{
		\centering
		\begin{tikzpicture}[scale=0.83, every node/.style={transform shape}]
		\node[shape=circle,draw=black,align=center,fill=red!7]  (A) at (8.8,0) {$A$};
		\node[shape=rectangle,draw=black,align=center] (v1) at (5.8,1.7) {$VM_1$};
		
		\node[shape=rectangle,draw=black,align=center] (v2) at (7.2,-2) {$VM_2$};
		
		\node[shape=rectangle,draw=black,align=center] (v3) at (4,0) {$VM_3$};
		
		\node[shape=rectangle,draw=black,align=center] (v4) at (2.5,-2.3) {$VM_4$};
		
		\node[shape=rectangle,draw=black,align=center] (v5) at (1,1) {$VM_5$};
		
		\node[shape=circle,draw=black,align=center,fill=red!7] (DB) at (-0.6,-0.4) {$DB$};
		
		\path [->] (A) edge node[left] {} (v1);
		\path [->] (A) edge node[left] {} (v2);
		\path [->] (v1) edge node[left] {} (v3);
		\path [->] (v2) edge node[left] {} (v3);
		\path [->] (v2) edge node[left] {} (v4);
		\path [->] (v3) edge node[left] {} (v5);
		\path [->] (v3) edge node[left] {} (v4);
		\path [->] (v4) edge node[left] {} (v5);
		\path [->] (v5) edge node[left] {} (DB);
		\node[shape=or gate US,thick,rotate=90,draw=black] (at4) at (2.5,-4.3) {OR}
		child
		{
			node{$\dots$}
		};

		\node[shape=or gate US,thick,rotate=90,draw=black] (at1) at (5.8,-4.3) {OR}
		child
		{
			node{$\dots$}
		};
		
		\node[shape=or gate US,thick,rotate=90,draw=black] (atdb) at (-0.6,-4.3) {OR}
		child
		{
			node{$\dots$}
		};
		\tikzset{grow'=down}
		\tikzset{every tree node/.style={anchor=base west}}
		\tikzstyle{level 1}=[sibling distance=7mm]
		\node[shape=or gate US,thick,rotate=90,draw=black] (at10) at (1,-3.9) {OR}
		child
		{
			child
			{
				node{$v_n$}
			}
			child
			{
				node{$\dots$}
			}
			child
			{
				node{$v_1$}
			}
			child
			{
				node{$v_0$}
			}
		};

		\node[shape=or gate US,thick,rotate=90,draw=black] (at6) at (4,-3.9) {OR}
		child
		{
			child
			{
				node{$v_n$}
			}
			child
			{
				node{$\dots$}
			}
			child
			{
				node{$v_1$}
			}
			child
			{
				node{$v_0$}
			}
		};

		\node[shape=or gate US,thick,rotate=90,draw=black] (at2) at (7.2,-3.9) {OR}
		child
		{
			child
			{
				node{$v_n$}
			}
			child
			{
				node{$\dots$}
			}
			child
			{
				node{$v_1$}
			}
			child
			{
				node{$v_0$}
			}
		};
		
		\path [-] [dashed] (v5) edge node[left] {} (at10);
		\path [-] [dashed] (v4) edge node[left] {} (at4);
		\path [-] [dashed] (v2) edge node[left] {} (at2);
		\path [-] [dashed] (v1) edge node[left] {} (at1);
		\path [-] [dashed] (v3) edge node[left] {} (at6);
		\path [-] [dashed] (DB) edge node[left] {} (atdb);
		
		\end{tikzpicture}
		}
		\caption{The cloud model example~\cite{hong2012harms} showing an attacker ($A$) and a target (e.g., database, or DB) where (a) illustrates the cloud model and (b) shows the two-layer HARM of the cloud example including an AG in the upper layer and ATs in the lower layer capturing vulnerabilities (denoted by $v_i$) existing on each VM.}
		\label{fig:HARM}
\end{figure}
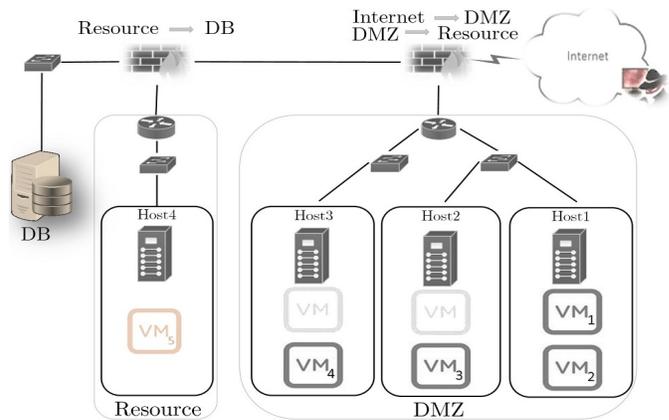
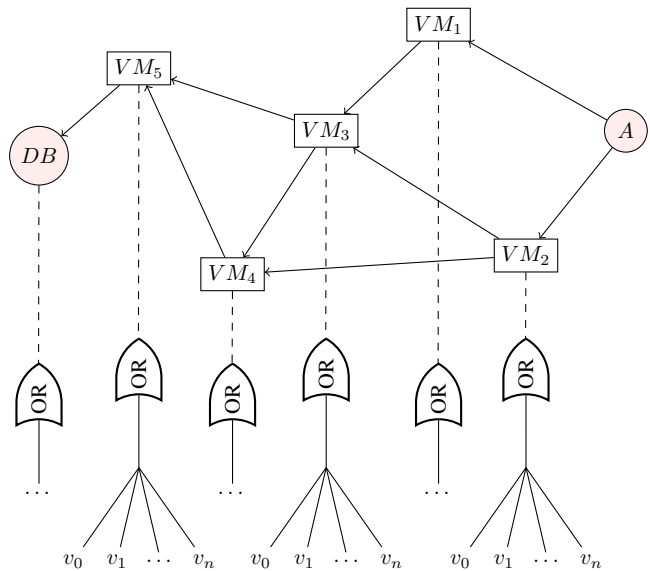

\subsubsection{Graphical Security Model-based MTD Evaluation}
The graph-based security models (GSM) have been proposed and widely used in modeling and analysis of the network security using attack graphs (AGs)~\cite{Sheyner:2002AG} and attack trees (ATs)~\cite{saini2008threat}. The hierarchical attack representation model (HARM)~\cite{hong2012harms} has been developed to model a system's security features with two layers, an upper layer and a lower layer. The upper layer represents a network's reachability information (i.e., network topological information) using an AG while the lower layer represents a node's vulnerability information using ATs, as shown in Fig.~\ref{fig:HARM}. A comprehensive assessment of the effectiveness of MTD techniques (i.e., shuffling, diversity, and redundancy) have been developed using the HARM~\cite{Alavizadeh18, Hong:ScalableAssMTD2014, hong2016assessing}. \citet{hong2018dynamic} proposed a technique to capture dynamic changes in the network resulted from deploying MTD techniques through the simulation. To this end, they incorporated MTD techniques into the temporal graph-based graphical security model~\cite{yusuf2016security} and then evaluated their MTD techniques using the dynamic security metrics.

\vspace{1mm}
\noindent {\bf Pros and Cons:} The main advantages of using GSMs are the ease of evaluation and representation. Furthermore, GSMs can be adopted to compute various security metrics based on the MTD application. A GSM can be easily visualized and help the network administrators or cloud providers to find out the vulnerabilities of the network and choose appropriate defensive strategies like MTD techniques. However, generating and analyzing GSMs in large-scale networks suffers from a scalability issue. Although this has been improved by using HARM which is more scalable than the other GSMs, it is still challenging to model the very large-scale networks using GSMs.

\begin{table*}[th!]
    \caption{Evaluation Methods of MTD Techniques: Pros vs. Cons.}
    \vspace{-2mm}
    \label{tab:eval-methods}
    \centering
    \begin{tabular}{|P{2.5cm}|P{7cm}|P{7cm}|}
    \hline
       {\bf Evaluation Method}  & {\bf Pros} & {\bf Cons} \\ \hline
        Analytical Models & Providing theoretical background and understanding with minimum cost; ease of modeling, representation, and evaluation & Scalability issues with some analyaitcal models (e.g., GSM, SPN); limitations in reflecting realistic scenarios due to abstraction/simplicity traded off for efficiency \\ \hline
        Simulation & Better flexibility in attack/system modeling than analytical models; easy parameterization for sensitivity analysis & Limitations due to inherent uncertainty toward real-world applications \\ \hline
        Emulation & High validity; high modeling flexibility; more realistic scenarios & Lack of scalability due to issues in limited hardware computational resources; still maintaining a certain level of abstraction/simplicity  \\ \hline
        Real Testbeds & Highest flexibility and validity in modeling and experiments among all other available evaluation tools & Least scalability in complexity and cost among all available evaluation tools \\ \hline
    \end{tabular}
\end{table*}

\subsection{Simulation Model-based MTD Evaluation}
Most studies evaluating the performance of MTD approaches have been validated based on simulation experiments. Compared to analytical models, simulation models have more flexibility in modeling specific attack behaviors (e.g., periodic attacks~\cite{jia2013motag}, DDoS attacks~\cite{zhuang2012simulation}, Multi-Armed Bandit policy-based attacks~\cite{penner2017combating}, sequential attacks on attack surface~\cite{Casola2013IRI}), and various testing scenarios characterizing unique environmental features (e.g., cloud-based service computing ~\cite{peng2014moving}, wireless sensor networks using IPv6-based MTD~\cite{Zeitz2017IoTDI}). Some simulation studies add MTD features into existing network simulators (e.g., {\tt NeSSi2}~\cite{zhuang2012simulation}).

\vspace{1mm}
\noindent {\bf Pros and Cons:} Simulation models provide high flexibility in modeling diverse types of attacks, environmental conditions, and/or different types of MTD techniques as most design features can be easily parameterized without much restriction. Due to the flexibility and experimental capability to conduct sensitivity analysis by varying the values of key design parameters, it allows us to easily obtain meaningful insights before performing the implementation in a real system. However, the simulation model has its inherent limitations because all possible variables may not be captured due to the nature of existing, uncontrollable uncertainty. Therefore, the lessons obtained from the simulation studies may not be realized in real testbed-based experiments or systems.

\subsection{Emulation Model-based MTD Evaluation}
Compared to the simulation-based studies, results from emulation testbed-based experiments can even provide a higher validity on experimental results although emulation-based studies are not common as much as simulation studies. \citet{Aydeger:2016Mitigating} proposed an SDN-based route mutation technique to deal with DDoS attacks which are validated via the implementation on the {\em Mininet} emulator with a Floodlight SDN controller. \new{Further, \citet{aydeger2019moving} defined a route mutation MTD technique for the Internet Service Provider (ISP) network context through NFV and virtual shadows network aiming to thwart possible DDoS attack. Their route mutation method makes it difficult for the attackers to perform attack reconnaissance phase and obtain network topology information. They implemented their work in an emulated environment using Mininet and evaluated the effectiveness of their framework in terms of success rate and overhead which measures defensive costs, such as storage cost and end-to-end (ETE) delay.}
\citet{Jafarian:OFRHM2012} presented an OF random host mutation technique, and validated it via the {\em Mininet} emulator with an NOX SDN controller.  \citet{Skowyra:MTD'16} constructed an enterprise-like network topology using the {\em Mininet} emulator, and validated the network performance in terms of time-to-first-byte and total download time, and the security robustness is evaluated by testing several classes of cyberattacks.
 
\vspace{1mm}
\noindent {\bf Pros and Cons:} Emulation models can provide a higher validity than simulation models in the experimental results validated through them. In addition, they also provide higher flexibility in modeling realistic attack types and MTD techniques than simulation-based evaluation. However, most emulation-based test environments (e.g., {\em Mininet}) running on a single machine share the same hardware resources for all emulation elements, making it difficult to evaluate experiments for large-scale networks. 

\subsection{Real Testbed-based MTD Evaluation}

\subsubsection{SDN-based Testbeds}
\citet{MacFarland15} proposed a host-based MTD technique to defend against network reconnaissance attacks on an SDN environment and validated their proposed method via the implementation of an SDN controller and DNS/NAT (Network Address Translator) functionalities on the {\em Ubuntu} OS using Python scripts.
\citet{Hong:2016Optimal} presented a shuffling-based optimal network reconfiguration method on SDNs. They constructed a real SDN testbed consisting of SDN-enabled hardware switches and an SDN controller to carry out performance analysis when the proposed shuffling-based MTD technique is deployed on the network.
\citet{Wang:MTD'17} constructed a real enterprise-like SDN topology with Open vSwitch (OVS) and Ryu SDN controller to verify their proposed U-TRI (Unlinkability Through Random Identifier) MTD technique.

\subsubsection{Cloud or Web-based Environment}
\citet{kampanakis2014sdn} presented an AG-based MTD countermeasure selection method. They constructed a prototypical virtualized cloud computing platform on Openstack and validated the performance overhead of the proposed method on the Cloud-based testbed.
\citet{vikram2013nomad} proposed a novel MTD technique that defends against web bots which automatically send requests to remote web servers by hiding the correct name/ID parameter values of HTML elements. Security effectiveness and performance overhead of the proposed technique are tested on the web platform testbeds.
\citet{penner2017combating} developed a set of MTD techniques that randomize the locations of VMs in the cloud against Multi-Armed Bandit (MAB) policy-based attacks. The authors demonstrated the proposed defense strategies in an OpenStack-based cloud environment and evaluated their performance based on migration downtime and network traffic.

\vspace{1mm}
\noindent {\bf Pros and Cons:} Real testbeds can help verify the performance of MTD techniques under more realistic system environments. However, it is difficult to evaluate the MTD techniques in large-scale networks. Most real testbed-based studies have focused on evaluating the performance of MTD in small or mid-sized networks.

We summarize the pros and cons of each evaluation method discussed above in Table~\ref{tab:eval-methods}.

\section{Application Domains of MTD Techniques} \label{sec:application-domains}
MTD techniques have been deployed in diverse application domains. We discuss how MTD techniques have been applied in the following environments: (1) enterprise networks; (2) Internet-of-Things (IoT); (3) cyber-physical systems (CPS); (4) software-defined networks (SDNs); and (5) cloud-based Web services. In this section, we aim to answer what MTD techniques, attack behaviors, design methodologies / theories, and evaluation methods are used in existing MTD techniques developed under each domain.

\begin{figure}[t]
	\centering
	\includegraphics[width=\linewidth]{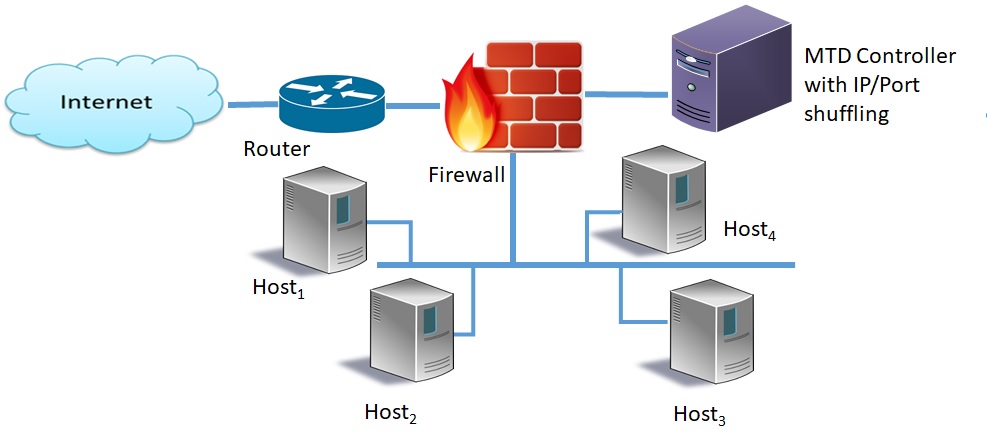}
	\caption{An example deployment of MTD techniques in an enterprise network~\cite{zhuang2013investigating}.}
	\label{fig:deployment_mtd_enterprise}
\end{figure}

\subsection{Enterprise Networks}
An enterprise networked system is homogeneously and statically configured. Since the enterprise networked system is built to operate in a static configuration, the attackers are relatively easy to plan and launch successful attacks to penetrate into the system. In addition, the attacker can develop mechanisms to make attacks dynamic, aiming at defeating detection mechanisms. 

\vspace{1mm}
\noindent {\bf MTD Techniques:} The common MTD techniques used for enterprise networks are mostly shuffling or diversity-based. The examples include platform migrations or system diversity~\cite{ben2016attacker, Cho18, evans2011effectiveness, feng2017signaling}, server location migrations~\cite{Ge14, wright2016moving}, software stack diversity~\cite{Zaffarano:Quantative2015}, proxy shuffling~\cite{jia2013motag}, or IP mutation~\cite{clark2015game}.

\vspace{1mm}
\noindent {\bf Main Attacks:} Most MTD approaches developed for enterprise networks countermeasured worm attacks~\cite{Ge14}, DDoS ~\cite{Ge14, jia2013motag, wright2016moving}, abstracted attacks in an attack-defense game~\cite{feng2017signaling}, scanning attacks~\cite{ben2016attacker, Cho18, clark2015game}, APT attacks~\cite{Zaffarano:Quantative2015}, or more sophisticated, multi-stage attacks, including circumvention attacks, deputy attacks, entropy reducing attacks, probing attacks, and incremental attacks~\cite{evans2011effectiveness}.

\vspace{1mm}
\noindent {\bf Key Methodologies:} Since majority of existing MTD approaches have used game theoretic approaches, many MTD techniques for enterprise networks have used in an attack-defense game where the MTD techniques are used as defense strategies. The examples include utility-concerned,  incentive-compatible MTD~\cite{Ge14}, game theoretic approaches (Bayesian \new{Stackelberg} game~\cite{feng2017signaling}, empirical game-theoretic analysis~\cite{wright2016moving}, moving target game~\cite{jia2013motag}), or decision utility model~\cite{ben2016attacker}. Some MTD research focused on developing the assessment tools to measure the performance of MTD techniques, such as quantitative MTD assessment framework~\cite{Zaffarano:Quantative2015} or model-based probability models to estimate the effectiveness of MTD techniques~\cite{DeLoach2014, evans2011effectiveness}.

\vspace{1mm}
\noindent {\bf Evaluation Methods:} As seen in Section~\ref{sec: evaluation-methods}, most MTD approaches have used simulation-based experiments for the performance validation. The similar trends are observed in evaluating MTD techniques for enterprise networks , which are mainly validated based on simulation models~\cite{ben2016attacker, clark2015game, feng2017signaling, Ge14, jia2013motag, wright2016moving, Zaffarano:Quantative2015} while the use of a model-based probability/analytical models was not common~\cite{Cho18}.

Fig.~\ref{fig:deployment_mtd_enterprise} depicts the architecture of the MTD deployment in an enterprise network~\cite{zhuang2013investigating}. The MTD mechanism is implemented in the configuration manager which produces effective configurations with intelligent adaptations.   

\vspace{1mm}
\noindent {\bf Pros and Cons:} Many MTD approaches for enterprise networks have been developed using game theoretic approaches by modeling an attack-defense game. As discussed in Section~\ref{subsec:game-modeling-solution}, modeling and formulating the interactions between an attacker and defender and estimating their utilities can be easily formulated using the game theoretic approaches. However, due to a certain level of abstractions on specific attack / defense behaviors or strategies, how much those can be applicable in practice still remained unclear. Further, most works are validated based on analytical or simulation models, which inherently introduce a certain level of abstraction and/or omission in modeling the behaviors of a system and players (i.e., an attacker and defender), which inherently limits the applicability.

\begin{figure}[t]
	\centering
	\includegraphics[width=\linewidth]{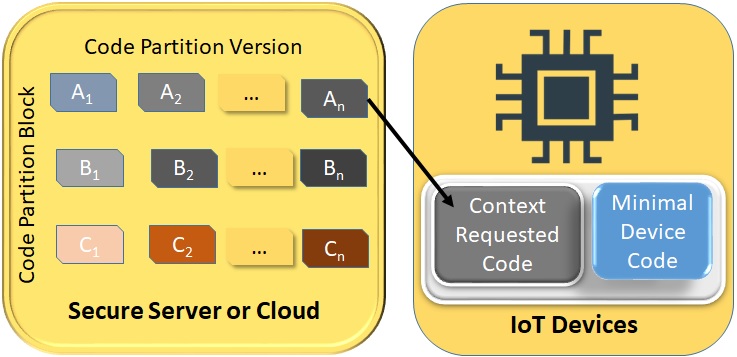}
	\caption{An example deployment of MTD in an IoT environment~\cite{Mahmood2016WFIOT}: Context-aware code partitioning and diversification.}
	\label{fig:deployment_mtd_iot}
\end{figure}

\subsection{Internet-of-Things (IoT)}
IoT refers to a network environment characterized by heterogeneous devices interacting with each other where the devices are controlled by machines and/or humans. Recently, the advance of IoT technologies has contributed to developing innovative applications in various domains~\cite{Roman13}. However, due to its large scale and severe resource constraints relying on limited bandwidth and/or power, conventional security / defense mechanisms (e.g., endpoint anti-virus software) has shown its limitations for applicability in IoT environments. Therefore, the threats and/or attacks encountered in IoT environments have been hurdles in providing seamless, normal services and operation of IoT-based systems~\cite{Ge17JNCA}. The MTD paradigm is an emerging, new technology concept that can provide the capability to protect an IoT system even under the challenges. In this section, we discuss the key MTD techniques, attacks, methodologies, and evaluation methods used to develop MTD technologies to protect IoT environments.

\vspace{1mm}
\noindent {\bf MTD Techniques:} In the literature, most IoT-based MTD techniques are shuffling or diversity-based. The examples include mutation of cryptosystem and/or firmware version~\cite{Casola2013IRI}, IP randomization~\cite{Sherburne2014CISR}, IPv6 rotation (i.e., $\mu$MT6D)~\cite{Zeitz2017IoTDI}, or code partitioning and diversification on IoT devices~\cite{Mahmood2016WFIOT}. Since IoT concerns severe resource constraints because most of them are mobile devices, developing lightweight shuffling-based MTD is a key concern in IoT-based MTD.

\vspace{1mm}
\noindent {\bf Main Attacks:} The most common attack behaviors considered in IoT-based MTD approaches are reconnaissance (scanning) attacks~\cite{Mahmood2016WFIOT, Sherburne2014CISR, Zeitz2017IoTDI, Zeitz18} although some conventional attacks, such as cryptographic attacks~\cite{Casola2013IRI}, are also considered. Although more intelligent, smarter attack behaviors have been extensively studied in IoT environments, IoT-based MTD techniques dealing with APT attacks have been rarely studied~\cite{Mohsin16}.

\vspace{1mm} 
\noindent {\bf Key Methodologies:} \citet{Casola2013IRI} identified reconfigurable architectural layers to develop shuffling-based or diversity-based MTD in terms of security and physical layers. IP shuffling-based MTD has been developed based on IPv6 for IoT devices 
~\cite{Sherburne2014CISR, Zeitz2017IoTDI, Zeitz18}. \citet{Mahmood2016WFIOT} used minimal trusted code in an IoT device, which can be erased after it was used in order to ensure security in a resource-constrained IoT environment.

\vspace{1mm}
\noindent {\bf Evaluation Methods:} As observed in other domains, simulation testeds are the most popular validation method. A real IoT testbed with end devices (e.g., laptops), called `6LoWPAN testbed'~\cite{Sherburne2014CISR}, is developed to validate IP shuffling MTD. Some recent studies developed the testbeds for IP mutation-based MTD using IPv6~\cite{Zeitz2017IoTDI, Zeitz18}. 

Fig.~\ref{fig:deployment_mtd_iot} shows an implementation of MTD in IoT devices where context-aware coding and diversification is used as an MTD technique where a minimal device code is used to ensure system security in a resource-constrained IoT environment~\cite{Mahmood2016WFIOT}.

\vspace{1mm}
\noindent {\bf Pros and Cons:}
The conventional security standard following the concept of defense-in-depth may not protect IoT devices because you cannot install those mechanisms into the IoT devices and difficult to view and monitor the activities happening inside the device.  An effective deployment of MTD technique in IoT that confuses an attacker at the very first phase of the cyber kill chain by APT attackers can make a target invisible, and invalidate their intelligence obtained from reconnaissance. Therefore, MTD makes the attackers harder to map the devices, exploit their vulnerabilities and launch the attacks. However, the constraints on IoT devices (e.g., CPU, energy consumption, memory) and network (e.g., low-bandwidth, high packet-loss) limit the effectiveness of MTD in IoT environments. 

\subsection{Cyber-Physical Systems (CPS)}
CPS is a system with cyber capabilities in physical worlds embracing humans, infrastructure or platforms that allow to communicate to each other~\cite{Poovendran10}. The advance of CPS has been made along with increased cyber capabilities in terms of communications, networking, sensing, and computing as well as enhanced capabilities of physical systems with materials, hardware, and/or sensors/actuators. The unique aspect of CPS is from the coexistence and coordinations between cyber capabilities and physical resources~\cite{Poovendran10}. In this section, we discuss what key MTD techniques, attacks, methodologies, and evaluation methods are considered in existing MTD techniques developed for CPS.

\vspace{1mm}
\noindent {\bf MTD Techniques:} MTD techniques have been developed to protect CPS environments. \citet{Li14} developed a lightweight, adaptive packet morphing technique; \citet{Potteiger2018HoTSOS} used an instruction set randomization to protect a vehicular network. Some MTD works are proposed to protect supervisory control and data acquisition (SCADA) system, such as IP hopping mutating the IP addresses of the gateway router in a power grid SCADA~\cite{Pappa2017ISGT}, Dynamic Generated Containment System (DGCS)~\cite{Chin16}, or IP hopping at an SCADA~\cite{Ulrich17}.

\vspace{1mm}
\noindent {\bf Main Attacks:} The attack behaviors considered in developing MTD techniques for CPS include code injection attacks in a vehicular network~\cite{Potteiger2018HoTSOS}, eavesdropping and/or traffic analysis attacks~\cite{Li14, Pappa2017ISGT}, IP scanning attacks~\cite{Pappa2017ISGT}, brute-force login, zero-day exploit, a malicious binary upload, and DDoS~\cite{Chin16}, IP spoofing, relay attack, vulnerability scanning, OS detection~\cite{Ulrich17}.

\vspace{1mm} 
\noindent {\bf Key Methodologies:} \citet{Li14} formulated a traffic morphing problem as an optimization problem aiming to minimize the number of redundant packets in the traffic morphing process. \citet{Pappa2017ISGT} developed an MTD architecture on Iowa State's PowerCyber testbed for targeted cyberattacks with a real world SCADA software and physical relays. \citet{Chin16} also developed the Dynamic Generated Containment System (DGCS) based on a virtualized Docker~\cite{Merkel14} to protect the host service and to detect a threat. \citet{Ulrich17} developed an IP hopping MTD based on four transformations of IP changes, two translations of the datagram's source address (SNAT) and the destination (DNAT), respectively.

\vspace{1mm}
\noindent {\bf Evaluation Methods:} \citet{Li14} conducted a simulation study to validate the proposed traffic morphing algorithm using a synthetic CPS network traffic and real-world network traces, including TCP and UDP packet-headers, to obtain target sessions. \citet{Pappa2017ISGT} validated an IP-hopping technique for a SCADA system on Iowa State's PowerCyber testbed. \citet{Chin16} used Global Environment
for Network Innovation (GENI), a virtual lab to conduct experiments in a large-scale network with multiple containers for validating the proposed MTD technique (i.e., Dynamic Generated Containment System, DGCS) based on the emulated system. \citet{Ulrich17} simulated a SCADA environment in terms of network traffics and delays to validate the performance of the IP hopping MTD in terms of throughput and delay.

\begin{figure}[t]
	\centering
	\includegraphics[width=\linewidth]{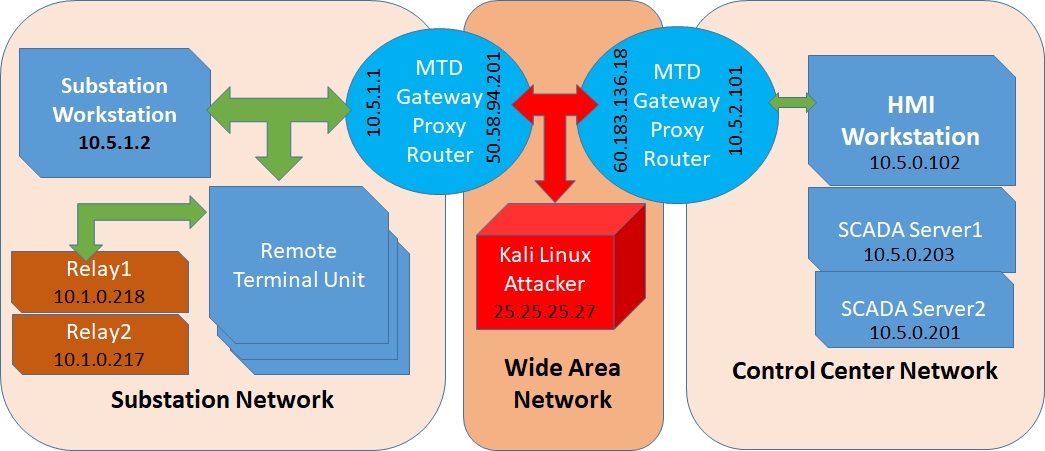}
	\caption{An example architecture of IP hopping MTD in an smart grid SCADA~\cite{Pappa2017ISGT}.
	}
	\label{fig:deployment_mtd_cps}
\end{figure}

Fig.~\ref{fig:deployment_mtd_cps} shows an example deployment of IP hopping based MTD in a smart grid SCADA network based on~\cite{Pappa2017ISGT}. 

\vspace{1mm}
\noindent {\bf Pros and Cons:}
SCADA-based systems are generally used for monitoring and controlling physical devices span in large geographic distances which are the part of national infrastructures (e.g., water distribution, oil and natural gas pipelines, power grids, and transportation systems) and are critical to a nation's economy and safety. MTD mechanisms to SCADA systems can protect the CPS system by adding an additional layer of defenses in addition to enhancing the effectiveness of MTD. However, an effective deployment of the MTD mechanisms should not adversely impact the core performance of a system, including safety, reliability, availability, and predictability in running operations of the CPS.

\subsection{Software Defined Networks (SDNs)} \label{subsec:sdn_app}
An SDN has emerged as a promising technology to decouple the network control plane from the data-forwarding plane for providing flexibility, robustness, and programmability to a networked system. In conventional networks, a routing algorithm in each switch makes packet forwarding decisions. By contrast, a controller on an SDN is designed to control the forwarding operations of the switches in a centralized manner. Thanks to this flexibility and programmability, the SDN technology has been leveraged by various cybersecurity network applications. In this section, we discuss the types of MTD techniques, attack behaviors considered, key methodologies, and evaluation methods considered and/or used for MTD technologies developed for SDN environments.

\vspace{1mm}
\noindent {\bf MTD Techniques:} IP shuffling / mutation based MTD is popularly used in SDN environments by leveraging OF switches and a centralized SDN controller~\cite{Jafarian:AddMutation2015, Jafarian:OFRHM2012, MacFarland15, Sharma18} where the SDN controller makes packet forwarding decisions at each OF switch and decisions on the updates of the flow tables at the switch. Network topology shuffling is also a well-known MTD technique~\cite{achleitner2016cyber, Aydeger:2016Mitigating} that has been applied in SDN environments in order to minimize security vulnerabilities in attack paths~\cite{hong2016assessing} by the SDN controller identifying an optimal network reconfiguration. Scalability issue in SDN environments is also managed by the SDN controller which creates scalable attack graphs~\cite{Chowdhary:2016SDN}. An SDN-based packet header randomization is proposed to realize unlinkability anonymity in communications where the SDN controller and OF switches take care of routing based on nonce information~\cite{Skowyra:MTD'16, Wang:MTD'17}.

\vspace{1mm}
\noindent {\bf Main Attacks:} Common attacks considered in SDN-based MTD approaches include reconnaissance (or scanning) attacks~\cite{Chowdhary:2016SDN, hong2013scalable, hong2016assessing, Jafarian:AddMutation2015, Jafarian:OFRHM2012, kampanakis2014sdn, Sharma18} and DDoS attacks~\cite{Aydeger:2016Mitigating, Steinberger:2018DDoS}, which can be countermeasured by using random IP mutation and/or network topology shuffling.

\vspace{1mm} 
\noindent {\bf Key Methodologies:} The key idea of deploying MTD techniques in SDN environments is to highly leverage its centralized structure with an SDN controller to optimize the configuration of the deployed MTD techniques, such as IP randomization / shuffling~\cite{Jafarian:AddMutation2015, Jafarian:OFRHM2012, MacFarland15, Sharma18, Steinberger:2018DDoS}, network routing paths~\cite{Hong:2016Optimal}, attack graphs / paths~\cite{Chowdhary:2016SDN}, port hopping~\cite{kampanakis2014sdn}, packet header randomization / obfuscation~\cite{Skowyra:MTD'16, Wang:MTD'17}, or virtual topology generation~\cite{Achleitner:2017Deceiving}. The key concerns in developing SDN-based MTD are resolving a scalability issue in an attack graph~\cite{Chowdhary:2016SDN} or IP shuffling~\cite{MacFarland15} and optimizing both security and performance in terms of minimizing security vulnerabilities while minimizing defense cost and service interruptions to users. 

\vspace{1mm}
\noindent {\bf Evaluation Methods:} Like other domains, simulation-based validation methods are popular; but some emulation-based testbeds are also developed for validating SDN-based MTD (e.g., a {\em Mininet} emulator with an NOX SDN controller~\cite{Jafarian:OFRHM2012}). \citet{Sharma18} used both probability-based and simulation models to validate the proposed IP multiplexing / demultiplexing MTD. \citet{Hong:2016Optimal} conducted security and performance analysis of their proposed network topology shuffling-based MTD in a real SDN testbed consisting of SDN-enabled hardware switches and an SDN controller. \citet{kampanakis2014sdn} showed the performance of the proposed SDN-based port hopping MTD based on Cisco's One Platform Kit. 

\begin{figure}[t]
	\centering
	\includegraphics[width=\linewidth]{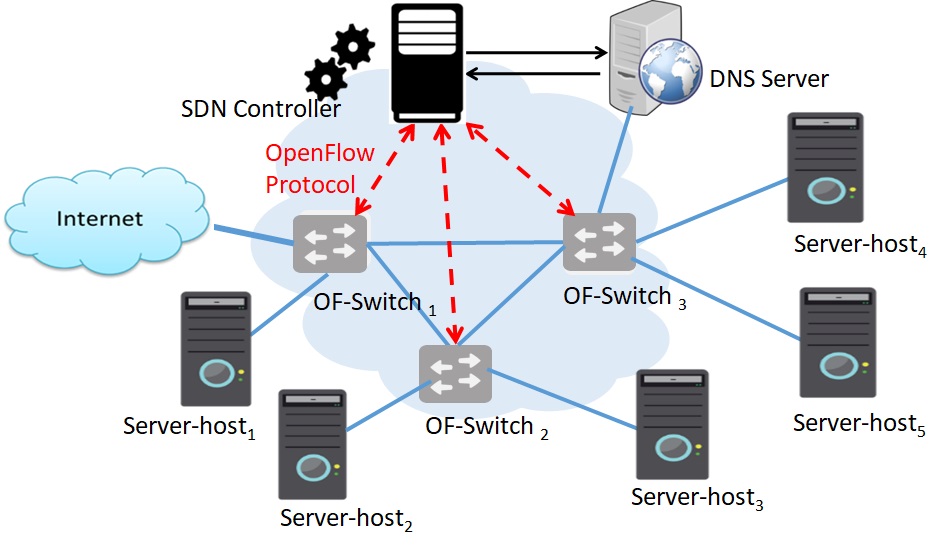}
	\caption{An example deployment of IP multiplexing / demultiplexing in an SDN environment~\cite{Sharma18}.}
	\label{fig:deployment_mtd_sdn}
\end{figure}

Fig.~\ref{fig:deployment_mtd_sdn} shows an example deployment of an SDN-based IP multiplexing / demultiplexing in which the SDN-controller implements the IP shuffling-based MTD mechanism. This is a centralized implementation of the SDN-based MTD at an SDN-controller in which the DNS response is intercepted by the controller which maps IP addresses and updates the flow tables at OF switches. 

\vspace{1mm}
\noindent {\bf Pros and Cons:} SDN is a promising technology that can provide flexibility, robustness, and programmability. This flexibility and programmability features of the SDN technology can be applied to several cybersecurity applications. The programmable interface afforded by the SDN environments can be comforted to implement proactive, adaptive defensive mechanisms, such as MTD techniques. The highly effective, scalable deployment of MTD can enhance system security and resiliency with multiple SDN controllers organizing centrally and hierarchically. The caveat is that introducing a hierarchical structure to maintain multiple SDN controllers may introduce extra overhead while it makes the SDN network highly scalable and can maximize the effectiveness of MTD. Therefore, finetuning the critical tradeoff between performance (e.g., defense cost and QoS) and security (e.g., security vulnerability) is a must.

\subsection{Cloud-based Web Services}
Cloud computing embraces both Internet-based applications provided as services such as hardware and system software in the data centers delivering those services~\cite{Armbrust10}. Could-based web services have been developed with the motivation of providing services based on centralized service management that enables providing professional, consistent, and uniformed quality services for all business unit services~\cite{Ghosh10}. However, the downside of the cloud-based centralized computing resources exposes a single point of risk and/or failure, such as `putting all your eggs in one basket' in which adversaries can easily target to attack. Further, maintaining a homogeneous computing infrastructure for uniform management of computing resources can also lead to a single vulnerability~\cite{Ghosh10}. MTD techniques have been explored to mitigate those security vulnerabilities and/or failure.

\new{\citet{Bardas:MTD-CBITS} proposed an MTD platform for cloud-based IT system and analyzed security benefits to verify its performance with an e-commerce scenario.  The authors studied how the MTD approach can be applied to the entire IT system and demonstrated its practicality in cloud-based environment using Automated eNterprise network COmpileR (ANCOR)~\cite{Unruh:USENIX2014}.  ANCOR is a framework for creating and managing the cloud-based IT systems. Their experimental evaluation showed that MTD systems managed and deployed using the proposed MTD approach increases attack effort/complexity.}

\vspace{1mm}
\noindent {\bf MTD Techniques:} The key MTD techniques addressed in the cloud computing domain include shuffling server replicas~\cite{Jia:CatchMeDSN2014}, VM migration~\cite{danev2011enabling, peng2014moving, zhang2012incentive}, VM snapshotting~\cite{peng2014moving}, web-application stack shuffling~\cite{sengupta2017game}, or platform diversity / migration~\cite{carter2014quantitative, carter2014game}. Since cloud-based web services take a centralized management, we can observe the MTD strategies are limited in changing configurations of system components, rather than varying network topologies which are mainly concerned in distributed environments. 

\vspace{1mm}
\noindent {\bf Main Attacks:} The common attacks considered in the existing cloud-based MTD approaches are DDoS~\cite{Jia:CatchMeDSN2014}, VM colocation attacks based on side channel attacks that can leak out private information of users~\cite{zhang2012incentive}, eavesdropping and message forgery / modification / dropping~\cite{danev2011enabling}, VM probing attacks~\cite{peng2014moving}, database / script / mainstream hackers~\cite{sengupta2017game}, or exploitation / data exfiltration attacks~\cite{carter2014game}. Since cloud-based web services may deal with a large volume of users' private information, it is noticeable that attacks leaking out the private information is also one of key concerns MTD needs to counteract.

\vspace{1mm} 
\noindent {\bf Key Methodologies:} 
\new{\citet{Bangalore:SCIT} proposed the Self Cleansing Intrusion Tolerance (SCIT), which is a new concept for securing servers based on a virtualization technology to rotate  the servers aiming to increase complexity for attackers and reduce the possible losses due to detection and prevention errors. \citet{Bangalore:SCIT} further extended their to design a Cloud-based SCIT~\cite{Nguyen:Cloud-SCIT} scheme for security enhancement of the applications and the services deployed in the Cloud. This approach is for a recovery-based intrusion tolerance system (ITS) designed to leverage the characteristics of the cloud computing and inter-cloud services. \citet{Nguyen:SCIT-MTD} also proposed an MTD-based self cleansing intrusion tolerance (SCIT-MTD) for securing a scalable web application services deployed in the cloud-based environments.} 

\citet{Jia:CatchMeDSN2014} proposed a replication method to make potential targeted servers as moving targets that can be isolated from adversaries. Further, the authors developed a client-server reassignment algorithm to mitigate interruptions of the service availability and quality.  \citet{zhang2012incentive} aimed to identify an optimal interval of VM migration in order to maximize security with minimum cost based on a game theory called Vickrey-Clarke-Groves (VCG) mechanism. \citet{danev2011enabling} proposed a key structure to allow secure migration of virtual Trusted Platform Modules (vTPMs) in private clouds. \citet{peng2014moving} formulated a service security model in clouds that can allow optimal configurations of VM migration / snapshotting and diversity / compatibility of migration. \citet{sengupta2017game} adopted a repeated Bayesian Stackelberg game to model a web-application stack shuffling MTD.   \citet{carter2014game} used a leader-follower game to identify an optimal use of strategies in choosing the right platform where their goal is to show that deterministic strategies developed based on statistical analysis outperforms simple random strategies.

\vspace{1mm}
\noindent {\bf Evaluation Methods:} Like the validation trends observed in other domains, most cloud-based MTD approaches are also evaluated based on simulation and analytical models~\cite{Jia:CatchMeDSN2014, peng2014moving, zhang2012incentive}, concerning the effectiveness of MTD in terms of security and cost. \citet{Jia:CatchMeDSN2014} used both simulation-based and prototype-based evaluations (i.e., using Amazon EC) to validate server shuffling-based MTD on clouds. \citet{danev2011enabling} integrated their proposed VM-vTPM migration MTD into a Xen hypervisor for the performance and security analysis. \citet{peng2014moving} conducted simulation-based performance analysis under various conditions of attack surface and service deployments. \citet{carter2014quantitative, carter2014game} conducted a simulation study in order to investigate the effect of game strategies in choosing platform migration on system security (i.e., mean time to compromise or MTTC). \citet{sengupta2017game} developed a real world MTD web application that can convert from an application in Python to one in PHP and from a MySQL database to a PostgreSQL database, or vice versa; and they identified an optimal solution based on a real testbed.

\begin{figure}[t]
	\centering
	\includegraphics[width=\linewidth]{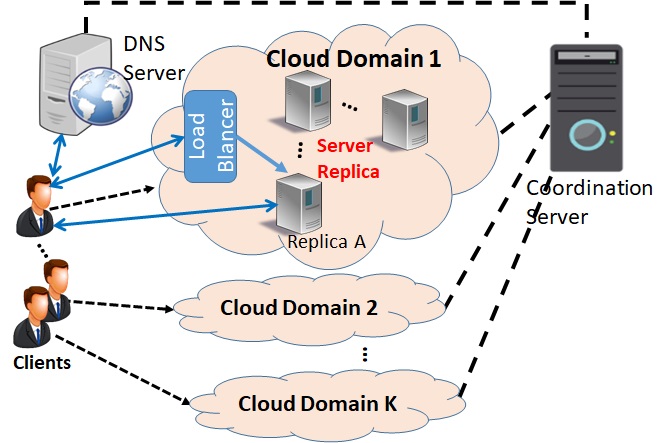}
	\caption{An example deployment of a cloud-based server-shuffling MTD~\cite{Jia:CatchMeDSN2014}.}
	\label{fig:deployment_mtd_cloud}
\end{figure}

Fig.~\ref{fig:deployment_mtd_cloud} describes an example implementation architecture of MTD for a cloud's services in which server-replica of the cloud-domain are shuffled~\cite{Jia:CatchMeDSN2014}.

\vspace{1mm}
\noindent {\bf Pros and Cons:} The newer cloud technologies have embraced the concept of infrastructure as a code; and an effective deployment of MTD techniques in clouds not only strengthens security, but also encourages innovation that leverages a new way of handling and managing dynamic configurations and infrastructure. With those technologies being leveraged, deploying the MTD techniques in clouds are more feasible for smaller and medium business houses as well as big enterprises and government. Therefore, highly scalable and cost-effective MTD deployment is allowed. However, it is inevitable to face challenges to make a good balance between security and performance in terms of minimizing security vulnerabilities and defense cost while maximizing service availability.

\section{Limitations} \label{sec:limitations}
In this section, we summarize the limitations we realized through the comprehensive survey conducted in this work as follows:
\begin{itemize}
\item \new{{\bf Limited investigations on the interplay between MTD and other defense mechanisms}: We have discussed event-based MTDs in Section~\ref{sec:mtd_classification} that describes the use of other defense mechanisms (i.e., IDS and/or IPS) to determine when to trigger an MTD operation and change attack surface. However, an attacker can take advantage of a system with a static and fixed placement of the defense mechanisms. Although one of MTD's key roles is to assist other defense mechanisms through cooperating with them, few prior work have investigated the cooperative role~\cite{Sengupta:GameSec2018, Chowdhary:MTDforDetection2018,Venkatesan:MTD2016}.  These approaches tackled the problem using an MTD approach which dynamically changes the placement of the IDS. However, their scope is limited to the placement of the detectors in the network. The MTD can help deploy other security defense mechanisms (i.e., IDS, IPS, deception, firewall) by cooperating with them as another layer of defense. However, few prior works have investigated how MTD assists in enhancing security and/or reducing the defense cost.}
\item {\bf Few studies investigating the effect of MTD on reducing attacks beyond the 
    reconnaissance stage}: Most attack behaviors considered in the existing MTD works occur during the reconnaissance stage.  This implies that MTD mainly deals with outside attackers as its primary goal is to protect a system before they break into the system. This significantly limits the applicability of MTD techniques although MTD can enhance system security and performance by dealing with attackers beyond the reconnaissance stage (e.g., inside attackers). 
    \item {\bf Lack of investigating the optimal deployment of multiple, hybrid MTD techniques}: Hybrid MTD approaches combining more than one technique among shuffling, diversity, and redundancy~\cite{alavizadeh2017effective, Alavizadeh18, alavizadeh2018evaluation, gorbenko2009using} have been proposed to maximize security. However, how to optimally deploy multiple techniques in terms of minimizing defense cost and maximizing system security and service availability has not been investigated.
    \item {\bf Metrics used in the existing MTD approaches have limitations in measuring multiple dimensions of a system's quality.} The state-of-the-art MTD metrics mostly only measure the effectiveness in terms of system security. A few studies have tried to assess performability aspects of the system but left out economical aspects (e.g., operational and/or capital cost). We discuss how the three metrics (i.e., security, performability, and economical costs) can be considered in Section~\ref{sec:insights-future-work}.  
    \item {\bf Lack of realistic testbeds}: Most MTD approaches have been validated based on analytical and/or simulation models while emulation-based or real testbed-based evaluation methods are rarely used. This indicates a critical need for developing better experimental testbeds for the verification and validation (V\&V) to test the performance of diverse types of MTD techniques. 
\end{itemize}


\section{\new{Conclusions}}
\label{sec:insights-future-work}

\new{In this paper, we conducted a comprehensive survey on MTD techniques, their key classifications, their key design dimensions, common attack behaviors handled by the existing MTD approaches, and application domains considered in the MTD literature. For the future research, we summarized the insights and lessons obtained from this survey paper and addressed the future research directions in the MTD research.}

\subsection{Insights \& Lessons Learned}
We obtained the following insights and lessons from this study:
\begin{itemize}
    \item {\bf Proactive, adaptive, and affordable defense}: MTD takes a non-conventional perspective toward security, which aims to enhance security by changing the attack surface, rather than eliminating all the vulnerabilities of system components, representing the conventional security goal. Based on this philosophy, MTD can provide a way to build proactive, adaptive, and affordable defense mechanisms that can leverage the existing technologies while enhancing system security. 
    \item {\bf Increased synergy in cooperation with other defense mechanisms}: By adding another layer of defense that can proactively thwart potential attackers or hold inside attackers' actions, MTD can help an IDS detect intrusions more effectively while providing an alternative defense strategy to deal with vulnerabilities when deception is detected by an attacker.
    \item {\bf Importance of balancing multiple objectives}: Although MTD introduces another layer of defense that can enhance security, it also generates overhead and may interrupt services to be provided to legitimate users. The thorough analysis of critical tradeoffs between multiple objectives is necessary to improve the state-of-the-art MTD technology.
    \item {\bf Customized MTD dealing with unique characteristics of different application platforms}: We noticed that different types of MTD have been developed to deal with the unique characteristics of different application domains in order to fully leverage the advanced legacy technologies to maximize the effectiveness and efficiency of MTD.
    \item {\bf Diverse solution techniques for developing MTD}: Although game theoretic MTD approaches are dominant among all, other optimization techniques based on genetic algorithms or machine learning have been explored. 
    \item {\bf Validation of the performance of MTD with diverse metrics}: The effectiveness and efficiency of MTD are measured by using a variety of metrics in terms of the perspectives of both an attacker and defender. 
    \item {\bf Effective and efficient evaluation methods developed to ensure the verification and validation (V\&V) of MTD}: Most MTD approaches have been validated based on simulation or analytical model-based experiments. But some existing approaches conducted experiments under more realistic environments using emulation or real-testbed based evaluation methods.
\end{itemize}
Although we could obtain the insights and lessons from the comprehensive survey conducted in this work, we felt the future research should be geared toward the right direction. We discuss the future research directions as below.
\begin{figure}[t]
	\centering
	\includegraphics[height=7.5cm]{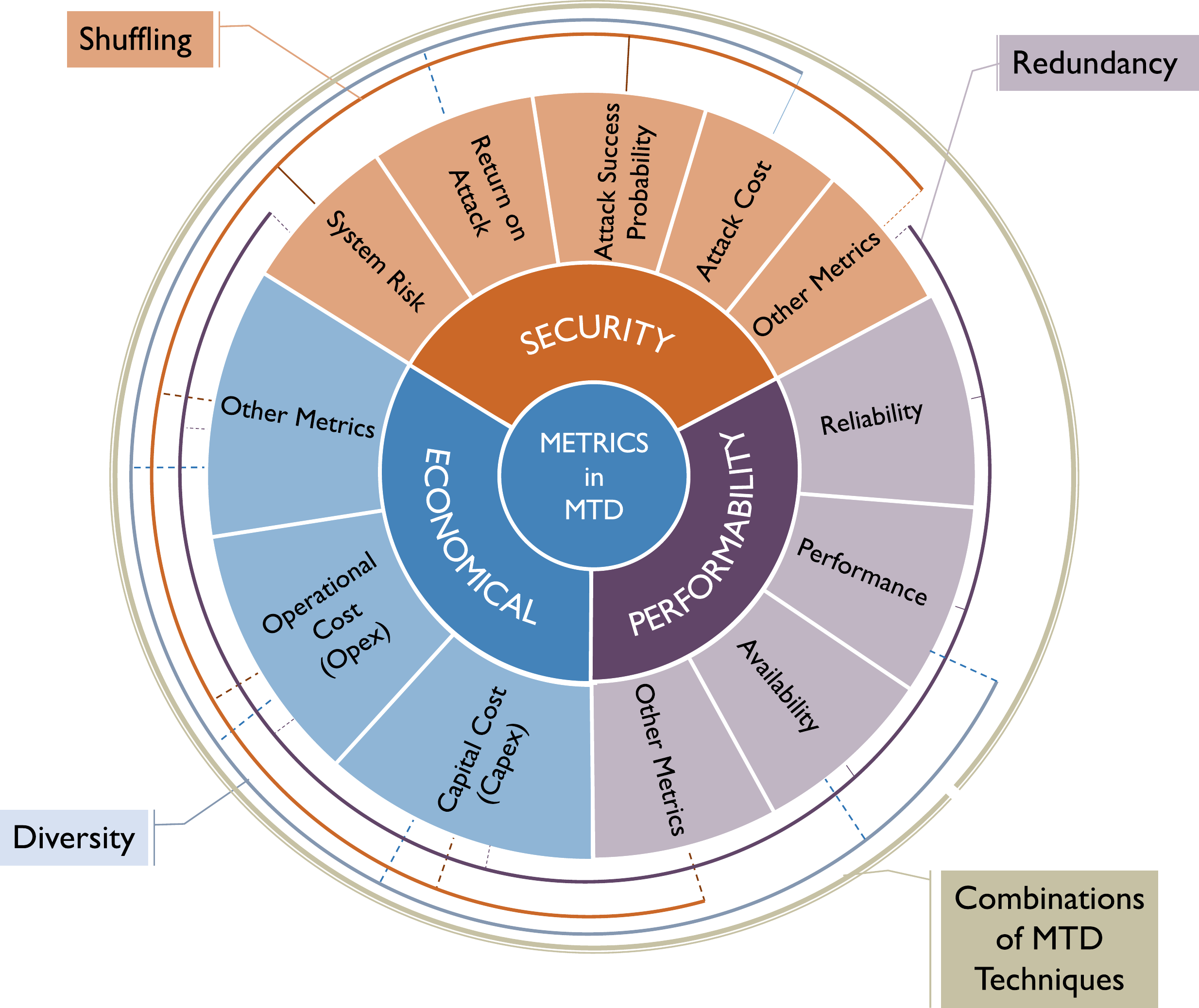}
	\caption{Security, performability, and economical MTD metrics.}
	\label{fig:metrics}
\end{figure}

\subsection{Future Research Directions}
We suggest the following future research directions in the MTD research domain:
\begin{itemize}
    \item {\bf A better MTD classification needs to be developed.} Although we used the operation-based MTD classification based on shuffling, diversity, and redundancy, it only captures `how-to-move.' It does not capture other criteria such as `when to move' or `what to move.' We need to develop an MTD classification that can comprehensively embrace multi-dimensions of MTD properties which can provide a better understanding of key MTD operations.
    \item {\bf More adaptive MTD mechanisms need to be developed.} In the state-of-the-art MTD technology, the concepts and techniques of adaptive MTD are still immature.  The adaptivity in triggering MTD operations can be achieved based on the level of system vulnerabilities or attack patterns / strength, which requires the advanced detection or learning capability of the defender. How to learn an attacker's action and/or system security conditions is vital for the defender to make decisions for the optimal MTD deployment.
    \item {\bf Lightweight MTD should be developed with high granularity of meeting the needs for highly contested environments.} MTD can be a good defense solution to protect military tactical environments, where resources are often severely restricted and communications between nodes are fully distributed, requiring a fully autonomous decision process. However, most MTD approaches addressed in the state-of-the-art do not provide highly lightweight, distributed solutions. Lightweight, secure MTD should be built to meet multiple, conflicting objectives of the contested, tactical environments.
    \item {\bf More useful metrics are needed to maintain service availability to users.} While MTD introduces enhanced security, it may also hinder service availability to legitimate, normal users. However, most metrics are focused on measuring system security or attack success. Since the system has multiple objectives to meet in terms of security, defense cost, and service availability, there should be a balance to develop meaningful, useful metrics that can capture the impact introduced to users in terms of service quality. 
    \item {\bf The effectiveness and efficiency of MTD should be measured by system metrics embracing security, performability, and economical costs introduced by the MTD.} As shown in Fig.~\ref{fig:metrics}, we can consider the three metrics (i.e., security, performability, and economical cost) to evaluate MTD techniques. Security metrics can measure the effectiveness of the MTD while performability metrics can estimate the efficiency of the MTD. Moreover, the economical costs introduced by using the MTD should be considered particularly when MTD is implemented in real systems.  Hence, finding critical tradeoffs between these three metrics should be investigated in the evaluation stage. 
\end{itemize}

\section*{Acknowledgement}
This work was partially supported by US Army Research, Development and Engineering Command (RDECOM) International Technology Center-Pacific (ITC-PAC) and U.S. Army Research Laboratory (US-ARL) under Cooperative Agreement, FA5209-18-P-0037. The views and conclusions contained in this document are those of the authors and should not be interpreted as representing the official policies, either expressed or implied, of RDECOM ITC-PAC, US-ARL, or the U.S. Government. The U.S. Government is authorized to reproduce and distribute reprints for Government purposes notwithstanding any copyright notation here on.

\bibliographystyle{IEEETranSN}
\bibliography{mtd-bib}

\newpage

\end{document}